\documentclass[fleqn,usenatbib,useAMS]{mnras}
\usepackage[dvipsnames]{xcolor}
\usepackage{tikz,hyperref}

\definecolor{lime}{HTML}{A6CE39}
\DeclareRobustCommand{\orcidicon}{
	\begin{tikzpicture}
	\draw[lime, fill=lime] (0,0) 
	circle [radius=0.13] 
	node[white] {{\fontfamily{qag}\selectfont \tiny ID}};
	\draw[white, fill=white] (-0.0625,0.095) 
	circle [radius=0.007];
	\end{tikzpicture}
	\hspace{-2mm}
}

\foreach \x in {A, ..., Z}{\expandafter\xdef\csname orcid\x\endcsname{\noexpand\href{https://orcid.org/\csname orcidauthor\x\endcsname}
			{\noexpand\orcidicon}}
}
\usepackage{newtxtext,newtxmath}

\usepackage{booktabs}
\usepackage[T1]{fontenc}
\usepackage{ae,aecompl}

\usepackage{graphicx}	% Including figure files
\usepackage{amsmath}	% Advanced maths commands
\usepackage{booktabs}
\usepackage{times,soul}
\input{epsf}
\title[Born-again PNe with an LTP event]{Formation and Evolution of [Wolf-Rayet] Planetary Nebulae through a Late Thermal Pulse}

\author[J.~B.~Rodr\'{i}guez-Gonz\'{a}lez et al.]{J.~B.~Rodr\'{i}guez-Gonz\'{a}lez\thanks{E-mail:\,janis@iaa.es}$^{1,2}\orcidA$, R.~Orozco-Duarte$^{1}\orcidB$, J.~A.~Toal\'{a}$^{1}\orcidC$, M.~M.~Miller Bertolami$^{3,4}\orcidD$ \newauthor H.~Todt$^{5}\orcidE$, M.~A.~Guerrero$^{2}\orcidF$, L.~Conmy$^{6}$ and R.~Kuiper$^{6}\orcidG$\\
$^{1}$Instituto de Radioastronom\'{i}a y Astrof\'{i}sica, Universidad Nacional Aut\'{o}noma de M\'{e}xico, 58089 Morelia, Michoac\'{a}n, Mexico\\
$^{2}$Instituto de Astrof\'{i}sica de Andaluc\'{i}a, IAA-CSIC, Glorieta de la Astronom\'{i}a S/N, Granada 18008, Spain\\
$^{3}$Instituto de Astrof\'{i}sica de La Plata, Consejo Nacional de Investigaciones Cient\'{i}ficas y T\'{e}cnicas, Avenida Centenario S/N, La Plata B1900FWA, Argentina\\
$^{4}$Facultad de Ciencias Astron\'{o}micas y Geof\'{i}sicas, Universidad Nacional de La Plata, Avenida Centenario S/N, La Plata B1900FWA, Argentina\\
$^{5}$Institut f\"{u}r Physik und Astronomie, Universit\"{a}t Potsdam, Karl-Liebknecht-Stra\ss e 24/25, D-14476 Potsdam, Germany \\
$^{6}$Fakult\"{a}t f\"{u}r Physik, Universit\"{a}t Duisburg-Essen,
Lotharstra\ss e 1, D-47057 Duisburg, Germany
}

\date{\today}%Accepted XXX. Received YYY; in original form ZZZ}

\pubyear{2025}

\begin{document}
\label{firstpage}
\pagerange{\pageref{firstpage}--\pageref{lastpage}}
\maketitle

%\newcommand{\mmmb}[1]{{\color{purple} #1}}
\begin{abstract}
We present radiation-hydrodynamical simulations of the formation of a born-again planetary nebula (PN) triggered by a late thermal pulse (LTP). 
The 2D radiation-hydrodynamic simulations, performed with the {\sc pluto} code, follow the stellar and wind properties resulted from a stellar evolution model calculated with the Modules for Experiments in Stellar Astrophysics ({\sc mesa}) code.
This {\sc mesa} model uses a mass-loss prescription tailored for H-deficient [Wolf-Rayet]([WR])-type winds during the post-LTP phase. 
Our stellar model reproduces the nearly complete depletion of H expected after an LTP event, while matching the observed abundances and spectral types of iconic [WR]-type central stars of PNe. The simulations show that the H-deficient LTP ejecta forms a transient double-shell structure which, after $\sim$1000 yr, becomes fully mixed with the H-rich PN. The H-deficient ejecta mass ($2.9\times10^{-6}$~M$_\odot$) is too small compared to the H-rich nebular mass of $\sim1.1$~M$_\odot$ ejected during the last phases of stellar evolution to leave a lasting imprint on the nebular abundances, predicting an H-rich PNe around a [WR] central star. 
The injection of LTP material into the PN drives turbulence, clump formation, and enhanced mixing, providing an explanation to the larger expansion velocities and larger turbulent nebular structures of PNe with [WR] central stars compared to those with H-rich central stars. 
These results provide robust support for the born-again scenario as the origin of H-deficient [WR] central stars within H-rich PNe.
\end{abstract}

\begin{keywords}
(ISM:) planetary nebulae: general --- (stars:) planetary nebulae: central stars --- stars: evolution --- stars: low-mass --- stars: winds, outflows
\end{keywords}

%%%%%%%%%%%%%%%%%%%%%%%%%%%%%%%%%%%%%%%%%%%%%%%%%%
%%%%%%%%%%%%%%%%% BODY OF PAPER %%%%%%%%%%%%%%%%%%

\section{Introduction}\label{introduction}
\label{sec:intro}

Planetary nebulae (PNe) are gaseous nebulae formed as the result of the evolution of low- and intermediate-mass stars (0.8~M$_\odot  \lesssim M_\mathrm{ZAMS} \lesssim 8.0$ M$_\odot$)\footnote{ZAMS stands for zero-age main-sequence.}. These stars lose more than half of their initial mass through slow and dense winds ($\varv_\mathrm{AGB} \approx 10$ km~s$^{-1}$, $\dot{M}_\mathrm{AGB} \lesssim 10^{-5}$ M$_\odot$ yr$^{-1}$) when evolving through the asymptotic giant branch (AGB) phase  \citep[e.g.,][]{Ramstedt2020,Wallstrom2025}. 

Such evolution will take the star into the post-AGB phase, reaching effective temperatures high enough to ionise the expelled material. Another consequence of the temperature increase is the development of a line-driven wind that can reach velocities of 500--4000~km~s$^{-1}$ \citep[e.g.,][]{Guerrero2013}. This fast wind sweeps up and compresses the ionised material, forming an inner rim. This relatively simple interacting wind model, initially proposed by \citet{Kwok1978}, has been the basis for interpreting the formation and evolution of PNe. This paradigm has been extensively developed through detailed 1D radiation-hydrodynamical simulations \citep[e.g.,][]{Marten1991,Frank1994,Villaver2002a,Villaver2002b,Perinotto2004,Schonberner2005}, with multi-dimensional numerical calculations allowing the exploration of additional physics \citep[e.g.,][and references therein]{Frank1993,GarciaSegura1997,GarciaSegura2025,Mellema1997,Mellema1994,Toala2014,Zou2020}.

According to \citet{Weidmann2020}, 10--30\% of PNe are estimated to host a H-deficient central star (CSPN), which display emission-line spectra dominated by He and C closely resembling those of massive Wolf-Rayet (WR) stars \citep{Crowther1998,Acker2003}. Although they are classified using the same spectral methodology as massive WR stars, a distinguishing convention using square brackets, i.e., [WR], has been adopted to denote the low-mass progenitors of these evolved stars \citep[][]{vanderHucht1981}. The most widely accepted formation scenario for a [WR] CSPN is the so-called {\it born-again} event in which the CSPN experiences a thermal pulse during its post-AGB evolution  \citep{Schonberner1979,Iben1983,MB2024}. 

A born-again event occurs when a post-AGB star experiences a final He-shell flash after leaving the AGB. He-shell flashes, or thermal pulses, are quasi-periodic thermal instabilities that develop in the He-burning shell as it becomes geometrically thin on the AGB \citep{2013KippBook}. Occasionally, such thermal instabilities can develop after the star has already left the AGB \citep[e.g.,][]{Herwig1999,Lawlor2003,MB2006}. The sudden energy release causes the remaining stellar envelope to expand and cool, driving the star back towards the AGB region of the HR diagram. At the same time, processed material from the intershell region, enriched in He, C, and O, can be brought to the stellar surface and can be ejected within the pre-existing PN. The subsequent evolution proceeds on relatively short timescales: the expanded envelope contracts again, the effective temperature rapidly increases, and the star re-enters the post-AGB evolutionary track while developing a new fast stellar wind.

Depending on the specific moment when the born-again event takes place, three broad categories are defined \citep[see recent review by][and references therein]{MB2024}: 
(i) an AGB final thermal pulse (AFTP), if it occurs just after leaving the AGB phase, 
(ii) a late thermal pulse (LTP), when it occurs on the post-AGB evolution in the Hertzsprung-Russell (HR) diagram, and 
(iii) a very late thermal pulse (VLTP), if the star experiences the thermal pulse when descending the white dwarf cooling track. 
Depending on the specific properties of the thermal pulse (AFTP, LTP, or VLTP), the final abundances of the CSPN might vary \citep[see, e.g.,][]{Todt2015,MB2024}. In addition to the ejection of H-deficient material inside its associated PN, some AFTP, LTP, and all VLTP born-again events are predicted to produce a [WR] CSPN \citep{Herwig1999,Guerrero2012,Guerrero2018,Toala2015}. 

To date, there are not many objects that have been detected to experience a born-again event. Those attributed to the evolution of a VLTP were reported for the PNe A~58 \citep[V605 Aql;][]{Clayton1997} and Sakurai's Object \citep[V~4334 Sgr;][]{Nakano1996} on 1919 and 1996, respectively. In both cases, H-deficient material is expanding inside their extended H-rich PNe as resolved with the Atacama Large Millimetre Array \citep[ALMA;][]{Tafoya2022,Tafoya2023}. More evolved cases of born-again PNe are those of A\,30, A\,78, and HuBi\,1, in which the post-born-again wind seems to be driving the kinematics of their H-deficient ejecta \citep{Fang2014,RechyGarcia2020}. On the other hand, 
FG\,Sge and Hen\,3-1357 (a.k.a.\ SAO\,244567) have been proposed to have experienced LTP events \citep[e.g.,][]{Jeffery2006,Reindl2017}.

Attempts to model the formation and evolution of born-again PNe are scarce. 1D simulations by \citet{Ary2023} and 2D simulations from \citet{Toala2021} were tailored to study the morpho-kinematics and nebular properties of HuBi~1. \citet{Fang2014} presented 2D radiation-hydrodynamic simulations to peer into the complex physics that experiences the H-deficient material ejected at low effective temperatures when interacting with the post-born-again wind and ionising photon flux in the cases of the most extended and evolved cases, A\,30 and A\,78. None of those works, however, considered the varying properties of the progenitor star as it evolves through the born-again event.

In this paper we use a stellar evolution model that experiences an LTP event to study its impact in the dynamics and evolution of the host PN and the production of a [WR] CSPN. The richness of the physical properties unveiled by the interaction of the H-deficient ejecta with the previous PN are studied through 2D radiation-hydrodynamical simulations.

The paper is organized as follows. In Section~\ref{sec:methods} we describe the details of our stellar evolution model and the characteristics of the hydrodynamical simulations. The results are described in Section~\ref{sec:results}. A discussion is presented in Section~\ref{sec:discussion} and, finally, the summary and conclusions are provided in Section~\ref{sec:summary}.

\section{Methods}
\label{sec:methods}

We compute the full evolutionary sequence of a low-mass star that experiences a born-again event, particularly an LTP. 
The stellar evolution model reports the time-dependent evolution of the stellar parameters (mass, effective temperature, luminosity, radius).  It consistently takes into account  
the mass-loss rate adopting appropriate 
approximations depending on the evolving phase. After the stellar evolution model is completed, it is then used to compute the evolution of the stellar wind velocity and ionising photon flux that are ultimately used as input for radiation-hydrodynamic simulations. This is the standard procedure used to model the formation and evolution of PNe and Wolf-Rayet nebulae around massive stars \citep[see][and references listed in Section~\ref{sec:intro}]{Villaver2002a,Villaver2002b,Perinotto2004,Toala2011,GarciaSegura1996,Freyer2006}.

\subsection{Stellar Evolution Model}
\label{sec:MESA}

We computed stellar evolution models with the Modules for Experiments in Stellar Astrophysics ({\sc mesa}) code \citep[version r15140;][]{paxton2011, paxton2013, paxton2015, paxton2018, paxton2019,Jermyn2023}. The code is collaborative and is publicly available at its GitHub project link\footnote{\url{https://github.com/projectmesa}}.

\begin{figure}
\begin{center}
\includegraphics[width=0.94\linewidth]{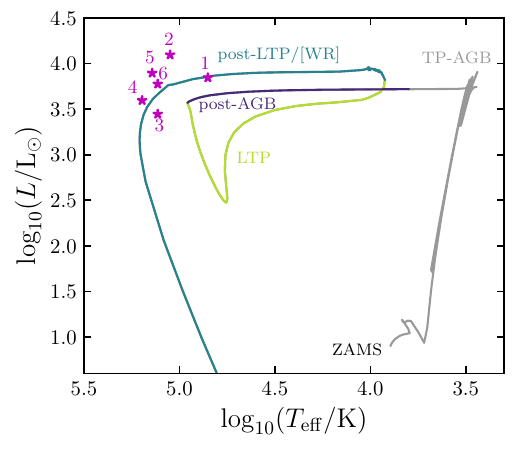}~
\end{center}
\caption{Evolution in the HR diagram of a $M_\mathrm{ZAMS}=1.7$ M$_\odot$ star model with $Z_\mathrm{ZAMS}$=0.02 that experiences an LTP event created with {\sc mesa}. Different segments illustrate the evolutive phases experienced by the star. The numbered star symbols correspond to the locii of the [WR] CSPN of NGC~40 (1), NGC~1501 (2), NGC~2371 (3), NGC~5189 (4), NGC~6905 (5), and PC~22 (6). }
\label{fig:HRD}
\end{figure}

We run a grid of stellar evolution models varying the initial mass ($M_\mathrm{ZAMS}$) and metallicity ($Z_\mathrm{ZAMS}$) to explore those that experience a born-again event. 
No rotation nor magnetic field are included in the present calculations. 
For the purposes of the present paper\footnote{The results of the stellar evolution grid will be presented in a subsequent paper (J.~B.~Rodr\'{i}guez-Gonz\'alez et al., in prep.) addressing the global properties of the sample and statistics of LTP events among post-AGB stars. }, we follow the evolution of an LTP event experienced by a star with $M_\mathrm{ZAMS} = 1.7$ M$_\odot$ and  $Z_\mathrm{ZAMS}= 0.02$ (see Fig.~\ref{fig:HRD}).  
In this model, the stellar mass at the onset of core He burning is 1.60648 M$_\odot$, which is drastically reduced by mass loss during the subsequent evolution to 0.56782 M$_\odot$ at the end of the AGB, and further reduced to 0.56777 M$_\odot$ after the LTP event. 
The total mass ejected during the born-again event is about $2.9\times10^{-6}$~M$_\odot$. The final mass of the star after descending the WD cooling track is 0.56623 M$_\odot$.

Previous studies showed that the efficiency of extra mixing plays a critical role in the evolution and surface composition of born-again stars \citep{Herwig1999, Blocker2001}. Using PG~1159 stars as observational constraints on the chemical composition of the intershell region, \citet{millerbertolami2016} and \citet{Wagstaff2020M} explored models with different efficiencies of convective boundary mixing (CBM) at the base of the pulse-driven convective zone (PDCZ). They found that increasing $f_{\rm PDCZ}$ enhances both C and O abundances in the intershell, improving the agreement with the observed PG~1159 abundances. Their analysis suggests that values in the range $f_{\rm PDCZ} \approx 0.004{-}0.014$ are required, with $f_{\rm PDCZ} \approx 0.008$ providing the best overall match to the typical observed C and O abundances. Models with $f_{\rm PDCZ} \lesssim 0.004$ show little change in intershell abundances and generally fail to reproduce the observed O enrichment. Following these results, overshooting efficiency in our models was taken as $f=0.014$ in the convective cores, both during H- and He-core burning.

When the star becomes H-deficient after the LTP, we adopt the opacity tables corresponding to R Coronae Borealis stars, based on {\sc mesa} calculations presented in \citet{Schwab2019}. These include the C and O enhanced opacities from OPAL \citep{Iglesias1993,Iglesias1996}, complemented with the \AE SOPUS CNO-enhanced abundances at lower temperatures \citep{Marigo2009}.

Fig.~\ref{fig:HRD} shows that our stellar evolution model exhibits the known behaviour of born-again event calculations \citep[see][]{Schonberner1979,Blocker2001,MB2005}.
 
\begin{figure}
\begin{center}
\includegraphics[width=0.9\linewidth]{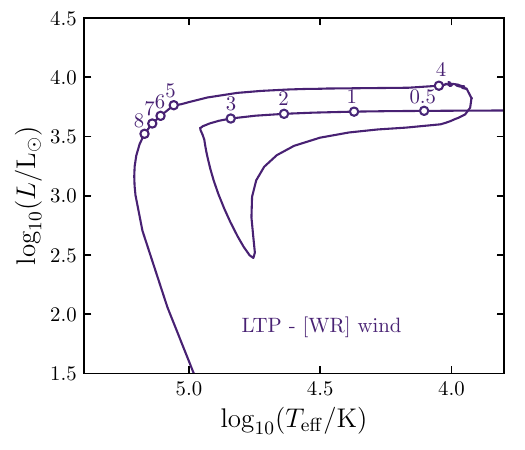}\\[-23pt]
\includegraphics[width=0.9\linewidth]{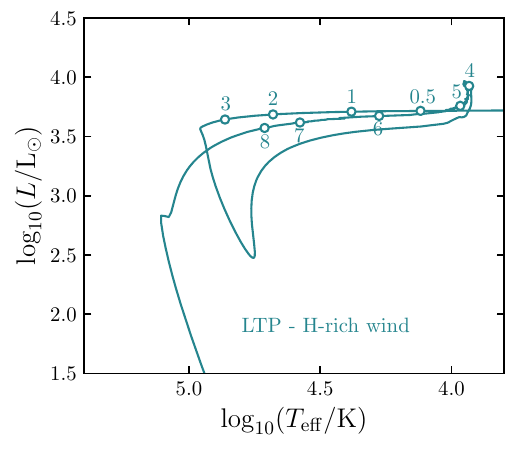}\\[-23pt]
\includegraphics[width=0.9\linewidth]{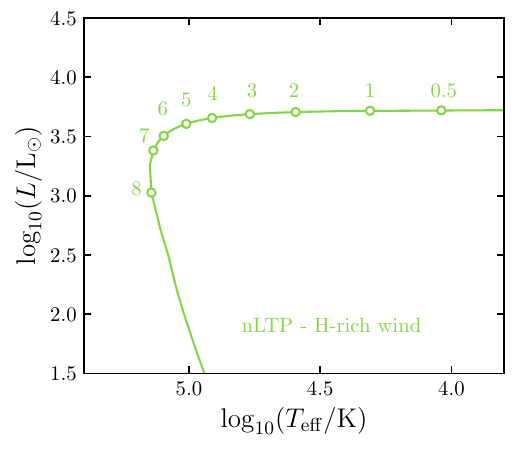}
\end{center}
\caption{A comparison of the post-AGB evolution between the three stellar evolution models analysed here. Top: The LTP model with a strong BA wind prescription (Eq.~\ref{eq:plane}). Middle: The LTP model with a feeble BA wind (Eq.~\ref{eq:MB}). Bottom: An evolving star that does not experience an LTP. The labelled open bullet symbols mark the 0.5, 1, 2, 3, 4, 5, 6, 7, and 8 kyr of post-AGB evolution.}
\label{fig:HRD2}
\end{figure}

The star evolves from the ZAMS phase to the upper region of the HR diagram, ultimately producing a thermally-pulsating AGB star (TP-AGB). We define the beginning of the post-AGB phase when the star increases its effective temperature to values larger than log$_{10}(T_\mathrm{eff}/\mathrm{K}) = 3.8$. The post-AGB phase duration until the star experiences the LTP event is about 3350 yr. The star is then taken back to log$_{10}(T_\mathrm{eff}/\mathrm{K}) < 4.0$ values. The duration of the LTP event in this model is 277 yr. The onset of the post-LTP phase is defined at the time when the H surface abundance by mass is reduced to 0.01, coincident with the maximum He surface abundance (see below). The post-LTP evolution is fast, reaching log$_{10}(T_\mathrm{eff}/\mathrm{K}) > 5.0$ in just 730 yr (see Fig.~\ref{fig:HRD}), ultimately taking the star towards the WD cooling track. The top panel of Fig.~\ref{fig:HRD2} illustrates the temporal evolution of our born-again stellar evolution model after the onset of the post-AGB phase.

\begin{figure*}
\begin{center}
\includegraphics[width=0.43\linewidth]{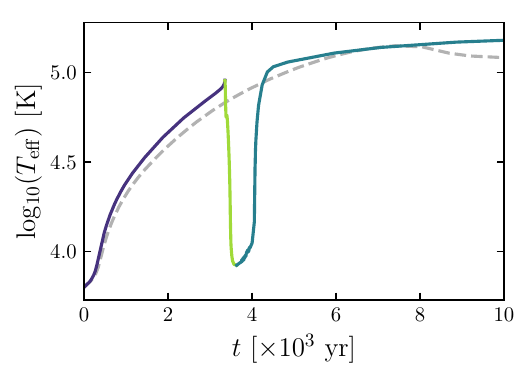}~
\includegraphics[width=0.43\linewidth]{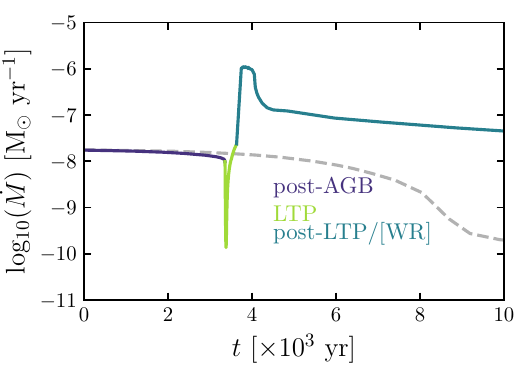}\\[-21pt]
\includegraphics[width=0.43\linewidth]{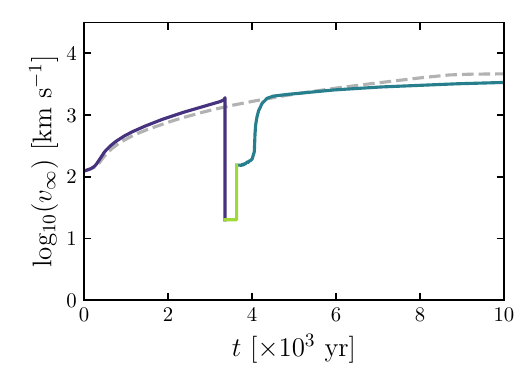}~
\includegraphics[width=0.43\linewidth]{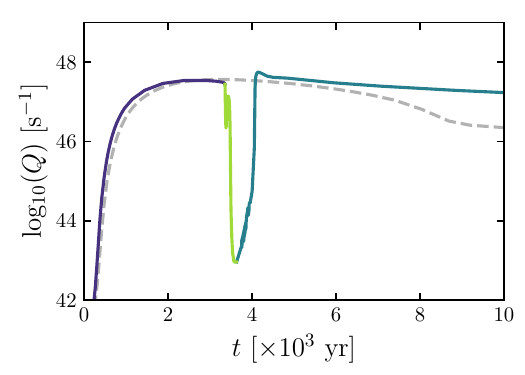}
\end{center}
\caption{Time evolution of the effective temperature ($T_\mathrm{eff}$ - top left), mass-loss rate ($\dot{M}$ - top right), stellar wind velocity ($\varv_\infty$ - bottom left), and ionising photon flux ($Q$ - bottom right) used in the radiation-hydrodynamical simulations. 
 The solid lines show the properties of the model that experiences the LTP event, with different colours denoting the post-AGB, LTP, and post-LTP phases. 
The dashed lines correspond to a classic post-AGB evolution. 
The time $t=0$ marks the onset of the post-AGB phase. }
\label{fig:postAGB}
\end{figure*}

Different mass-loss prescriptions suitable for each  evolutionary stage can be adopted for {\sc mesa} models. 
During the red giant branch we adopted the classical \citet{reimers1975} prescription with an efficiency parameter $\eta_\mathrm{R} = 0.5$, while during the AGB phase the formulation of \citet{blocker1995} using $\eta_\mathrm{B} = 0.2$ was used. 
For the post-AGB stage, we implement the prescription proposed by \citet{millerbertolami2016} where the mass-loss rate is calculated as 
\begin{equation}
    \left( \frac{\dot{M}}{\mathrm{M}_\odot~\mathrm{yr}^{-1}}\right) = 9.78 \times 10^{-15}  \left( \frac{L}{\mathrm{L}_\odot} \right)^{1.67}  \left( \frac{Z_0}{Z_\odot} \right)^{2/3},
\label{eq:MB}
\end{equation}  
\noindent with $L$ and $Z_0$ being the evolving stellar luminosity and initial surface metallicity, respectively. According to \citet{millerbertolami2016}, the mass-loss rate prescription of equation~(\ref{eq:MB}) is within a factor of two of those presented by \citet{Blocker1995_postAGB}.
It is worth emphasizing that the evolutionary timescale of the star at this phase is mostly determined by the H burning rate, with little dependence on the mass loss rate \citep[e.g.,][]{Paczynski1971,millerbertolami2016}.

After the star experiences the born-again event, it becomes an H-deficient [WR] CSPN. During this phase, we adopt the mass-loss rate prescription reported by \citet{Toala2024WR} for WR-type stars, given by 
\begin{equation}
\begin{aligned}
    \left(\frac{\dot{M}}{\mathrm{M}_\odot~\mathrm{yr}^{-1}}\right) = 7.76\times10^{-9} \left(\frac{L}{\mathrm{L}_\odot} \right)^{1.23}  
    \left(\frac{T_\mathrm{eff}}{\mathrm{K}} \right)^{-0.68}. 
    \label{eq:plane}
\end{aligned}   
\end{equation}
\noindent We call this prescription {\it strong Born Again wind}. The mass loss rate has a minor impact on the post-LTP evolution time scale, as the mass loss rate has negligible effects on the relatively large mass of the He-rich envelope.

\begin{figure}
\begin{center}
\includegraphics[width=1\linewidth]{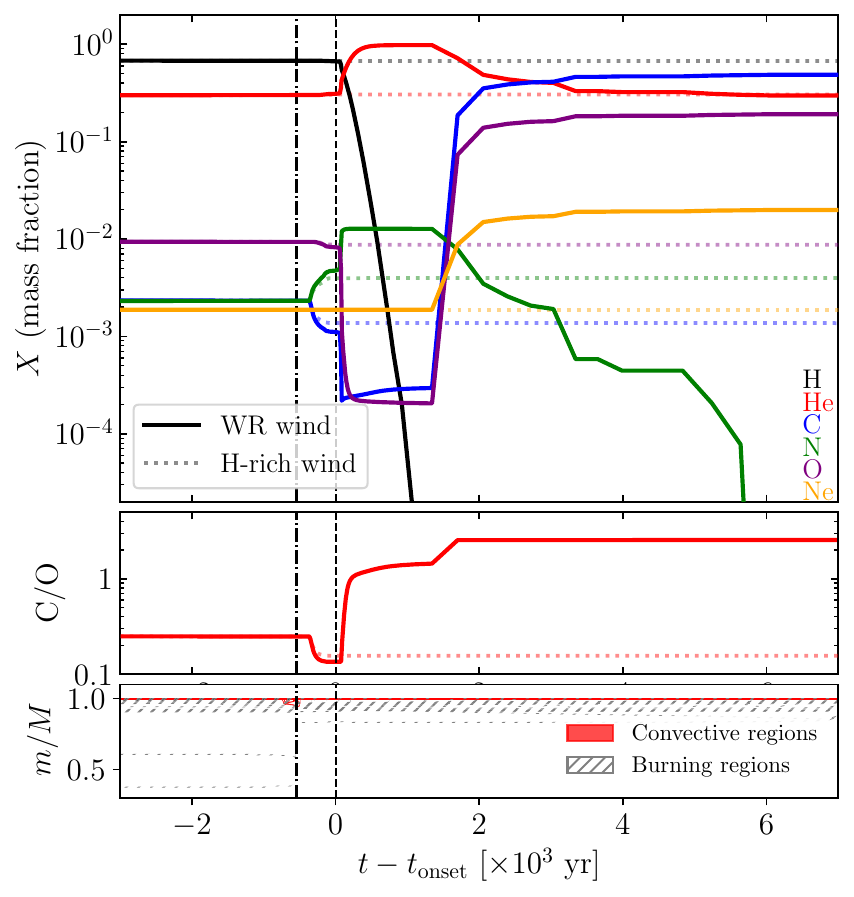}%\\[-18pt] %
\end{center}
\caption{ Evolution of the surface abundance (top panel), carbon-to-oxygen mass ratio (X$_{\rm C}/X_{\rm O}$, middle panel), and convective and burning zones (bottom panel). The solid curves in the upper and middle panels correspond to a the model computed with a [WR]-type wind prescription, while the dotted curves show the model with an feeble BA wind prescription. Different colors in the upper panel represent the surface abundances of H, He, C, N, O, and Ne. The bottom panel displays the Kippenhahn diagram of the strong Born Again wind model,%[WR]-wind model, 
where shaded and hatched regions denote convective and nuclear-burning zones, respectively. The vertical dashed line marks the onset of the post-LTP phase, whereas the dash-dotted line indicates the onset of the LTP event.}
\label{fig:M17_abund}
\end{figure}

\begin{table*}
\begin{center}
\caption{Surface elemental abundances by mass fraction previous and after the LTP event of the stellar evolution model used here compared to those predicted by \citet{MB2024} and six [WR]-type CSPNe from the literature$^{\dagger}$.}
\label{tab:analysis}
\setlength{\tabcolsep}{0.8\tabcolsep}
\begin{tabular}{lrccccccl}
\hline
    Model & Time$^\ddagger$ & H & He & C & N & O & Ne & Reference\\
      & (yr) & \\
\hline
Pre-LTP & $-1000$ 
& 0.683
& 0.301
& 2.35$\times10^{-3}$
& 2.33$\times10^{-3}$
& 9.41$\times10^{-3}$
& 1.88$\times10^{-3}$ & This work \\

Post-LTP & 2235 
& $<10^{-5}$
& 0.751
& 0.162
& 7.55$\times10^{-3}$
& 0.068
& 0.110 & This work\\

Post-LTP & 2800 
& $<10^{-5}$
& 0.550
& 0.311
& 4.27$\times10^{-3}$
& 0.122
& 0.013 & This work\\

Post-LTP & 3610 & $<10^{-5}$
& 0.420
& 0.403
& 2.04$\times10^{-3}$
& 0.159
& 0.017 & This work \\

Post-LTP & 6140 & $<10^{-5}$
& 0.310
& 0.481
& 3.51$\times10^{-6}$
& 0.189
& 0.019 & This work \\

\hline

Post-LTP & 
& $9\times10^{-3}$
& 0.42
& 0.43
& $1.5\times10^{-4}$
& 0.12
& 0.01 & \citet{MB2024}\\

Post-LTP & 
& 0.040
& 0.31
& 0.40
& $1.4\times10^{-3}$
& 0.20
& 0.022 & \citet{MB2005}\\

\hline
    Object & Spe. Type & H & He & C & N & O & Ne & Reference \\
\hline
NGC~40 & [WC8]
& $<0.02$
& $0.57\pm0.10$
& $0.40\pm0.10$
& $(7\pm4)\times10^{-4}$
& $0.03\pm2$
& $<0.3$ & \citet{Toala2019_ngc40}\\

NGC~1501 & [WO1]
& \dots
& $0.60\pm0.10$
& $0.30\pm0.10$
& \dots
& $0.15\pm0.05$
& \dots & \citet{Rubio2022}\\

NGC~2371 & [WO1]
& \dots
& 0.71
& 0.20
& $1.0\times10^{-3}$
& 0.06
& 0.03 & \citet{GG2020}\\

NGC~5189 & [WO1]
& \dots
& 0.58
& 0.25
& 0.01
& 0.12
& 0.04 & \citet{Keller2014}\\

NGC~6905 & [WO2]
& $<0.05$
& $0.55\pm0.05$
& $0.35\pm0.05$
& $<10^{-4}$
& $0.08\pm0.01$
& $0.02\pm0.01$ & \citet{GG2022}\\

PC~22 & [WO1]
& $<0.25$
& $0.76\pm0.10$
& $0.10\pm0.02$
& $<3\times10^{-3}$
& $0.10^{+0.04}_{-0.02}$
& $0.04^{+0.02}_{-0.01}$ & \citet{Sabin2022}\\
\hline
\end{tabular}
\begin{minipage}{0.9\textwidth}
\footnotesize
$^\dagger$Undetermined abundances are listed with dots.\\
$^\ddagger$Time $t=0$ yr corresponds to the onset of the post-LTP evolution. For the total PN age of the model presented here, 3350 yr has to be added.
\end{minipage}
\end{center}
\end{table*}

The {\sc mesa} stellar evolution calculations result in the detailed temporal evolution of the stellar parameters ($L$, $R$, and $T_\mathrm{eff}$). 
These stellar parameters are subsequently used to derive the mass-loss rate ($\dot{M}$), ionizing photon flux, and wind velocity 
throughout the different evolutionary phases. 
The mass-loss rate is derived within {\sc mesa} using the different prescriptions described above. 
Meanwhile the ionizing photon flux is computed assuming that
the star emits as a blackbody and integrating for wavelengths shorter than 912 \AA.
During the post-AGB phase and when moving towards the left in the HR diagram after the LTP event, the star develops line-driven winds with terminal wind velocities $\varv_{\infty}\approx1000$~km~s$^{-1}$ \citep{Lamers1999}. Here $\varv_\infty(t)$ has been estimated by adopting the analytical approximations presented in section 6 of \cite{Kudritzki1989}, which has been demonstrated to work for post-AGB stars \citep[e.g.,][]{Welch1999,Corradi2000,Perinotto2004,Toala2014,Orozco-Duarte2026}.

On the other hand, there is not much information on the wind properties during born-again events. Nevertheless, the densest clumps in the born-again PNe A\,30 and A\,78 exhibit expansion velocities as slow as $\approx$30 km~s$^{-1}$ \citep{Meaburn1996,Meaburn1998,Fang2014}, suggesting that the born-again event produced relatively slow velocities. For simplicity, we adopt a constant wind velocity of 20~km~s$^{-1}$ during the LTP event. We note, however, that the ejecta in born-again PNe seem to have bipolar ejections expanding at velocities of a few times 100~km~s$^{-1}$ \citep[e.g.,][]{RechyGarcia2020,RodriguezGonzalez2022,MontoroMolina2024}. 
Such asymmetrical high-velocity components might point at the presence of companions, an effect that is not modelled in the present work and that will be pursuit in subsequent studies. Nevertheless, we remark that \citet{Toala2021} studied the impact of different born-again ejecta velocities for the case of HuBi\,1 and noted that, while the kinematic ages of the H-deficient ejecta would be different, the density structures are very similar.

Fig.~\ref{fig:postAGB} presents the evolution with time of the effective temperature ($T_\mathrm{eff}$ -- top left), mass-loss rate ($\dot{M}$ -- top right) and stellar wind velocity ($\varv_\infty$ -- bottom left) before, during, and after the LTP event. In addition, the figure also shows the evolution of the ionising photon flux ($Q$ -- bottom right). 
Note that the ionisation flux during the LTP event is reduced considerably compared to other evolutionary phases as a natural consequence of the reduction of $T_\mathrm{eff}$ during this phase.

 Finally, Fig.~\ref{fig:M17_abund} presents the evolution of the surface abundances (top) before, during, and after the LTP event for the models computed with a [WR]-type wind prescription (solid line) and a model that does not experience a born-again event (dotted lines). 

The figure illustrates the rapid depletion of H and the corresponding enhancement of He at the onset of the LTP phase, particularly in the 
strong Born Again wind, which evolves toward an H-deficient surface composition.  Representative abundance sets corresponding to the pre-LTP stage and four different epochs during the post-LTP evolution are listed in Table~\ref{tab:analysis}. 

\subsection{Auxiliary Stellar Evolution models}

For further discussion, we use two extra stellar evolution models. The first one corresponds to a model experiencing an LTP event. It has exactly the same evolution as model shown in Fig.~\ref{fig:HRD}, but it adopts the wind mass-loss rate prescription defined for feeble BA wind (Eq.~\ref{eq:MB}) during the post-LTP phase. Its evolution is illustrated in the middle panel of Fig.~\ref{fig:HRD2}. This model is used to assess the impact of the mass-loss rate prescription for H-deficient WR winds. 

The second model, is that of a star that does not undergo a born-again event, which we have labelled as "nLTP". Its evolution is identical to that of the 1.7~M$_\odot$ model during the AGB and early post-AGB phases. However, this control model proceeds through the remainder of the post-AGB evolution and transitions directly onto the WD cooling track without experiencing any form of born-again event. This model will be used to assess the production of the turbulent medium inside the born-again PN using hydrodynamical simulations. Its evolution is depicted in the bottom panel of Fig.~\ref{fig:HRD2}. 

The post-AGB evolution of the $T_\mathrm{eff}$, $\dot{M}$, $\varv_\infty$, and $Q$ parameters of the nLTP model are also plotted in Fig.~\ref{fig:postAGB} with a dashed-line to allow the comparison with the model that experiences the born-again event (solid line). It is expected that such nLTP model would result in a classic PN as shown by previous numerical works (see references in Section~\ref{sec:intro}). 
These plots confirm small differences in the time evolution of $T_\mathrm{eff}$, $\varv_\infty$, and $Q$ between the LTP and nLTP models, besides the LTP event itself. 
Otherwise notable differences are unveiled in the evolution of the $\dot{M}$ after the LTP event (top right panel), which are caused by the strong mass-loss rate approximation adopted during that phase.

\subsection{Radiation-hydrodynamic simulations}
\label{sec:simulations}

Radiation-hydrodynamic simulations are performed by means of the {\sc pluto} code \citep{Mignone2007}\footnote{\url{https://plutocode.ph.unito.it/}} following the stellar evolution models described in Section~\ref{sec:MESA}. We run 2D simulations adopting a finite-volume formulation with a Godunov-type solver using a HLLC Rieman solver. We carry out axially symmetric (2D) simulations in spherical coordinates under the assumption of mid-plane symmetry.

The computational domain extends from the stellar position up to 0.6\,pc in the radial direction and covers a polar angle range from 0 to $\pi/2$, resolved with 2400$\times$1200 uniformly spaced cells. This configuration allow us to resolve the hydrodynamical instabilities developed at the wind-wind interaction region \citep[see, e.g.][]{Orozco-Duarte2026}. To assess numerical convergence, we performed an additional simulation with twice the spatial resolution. The resulting gas dynamics and the formation of clumps showed no significant differences, confirming that the adopted resolution is sufficient for the present study.

The hydrodynamic equations are solved assuming an ideal equation of state. Self-gravity is not included in the simulations, given that gas dynamics are dominated by the mechanical input from stellar wind and thermal pressure gradients, and the gravitational potential of the CSPN. The initial ambient medium is uniform, with a number density of $n_\mathrm{ISM}$ = 1\,cm$^{-3}$ and temperature $T_\mathrm{ISM}$ = 100\,K.

Stellar wind is injected through the innermost 10 radial zones. The density in these cells is prescribed according to the mass-loss rate and wind velocity obtained from the {\sc mesa} models (see Fig.~\ref{fig:postAGB}) as:
\begin{equation}\label{eqn:1}
    \rho(r,t) =  \frac{\dot{M}(t)}{4 \pi r^{2} \varv_\infty (t)}.
\end{equation}
\noindent This way, the evolving stellar parameters are directly driving the inner boundary conditions of the flow.

Radiative cooling is included in the energy equation through a tabulated cooling  
curve. We generated the cooling curve using the {\sc cloudy} photoionisation code \citep{Ferland2017}. The cooling curve accounts for the ionisation state and chemical composition of the gas under conditions representative of PNe. In the present work, we adopt a cooling function corresponding to standard PNe abundances, based on the compilation of \citet{Aller1983}. 
The cooling curve spans a temperature range from 10 up to 10$^{9}$\,K and includes contributions from collisional excitation, recombination, and free-free emission processes. At temperatures above 10$^{5}$\,K cooling is dominated by collisionally ionised heavy elements, while at lower temperature line cooling from singly ionised species becomes important. Further details can be found in \citet{Orozco-Duarte2026}.

In addition to radiative cooling, the effect of ionising radiation form the CSPN is included through the {\sc Sedna} module \citep{Kuiper2020}, which is coupled to {\sc pluto}. {\sc Sedna} combines ray-tracing of direct UV photons from a point source with a flux-limited diffusion treatment of diffuse radiation arising from direct recombinations into the hydrogen ground state. 
The ionisation balance is solved using a time-dependent rate equation formalism that includes photoionisation, radiative recombination, and collisional ionisation processes. {\sc Sedna} also supports both the on-the-spot approximation and a more complete treatment of the diffuse ionising field, making it well-suited for multidimensional applications involving sharp ionisation fronts or shadowed regions. In this work, we adopt the on-the-spot approximation for the treatment of diffuse ionising radiation.

\begin{figure*}
    \centering
\includegraphics[width=\linewidth]{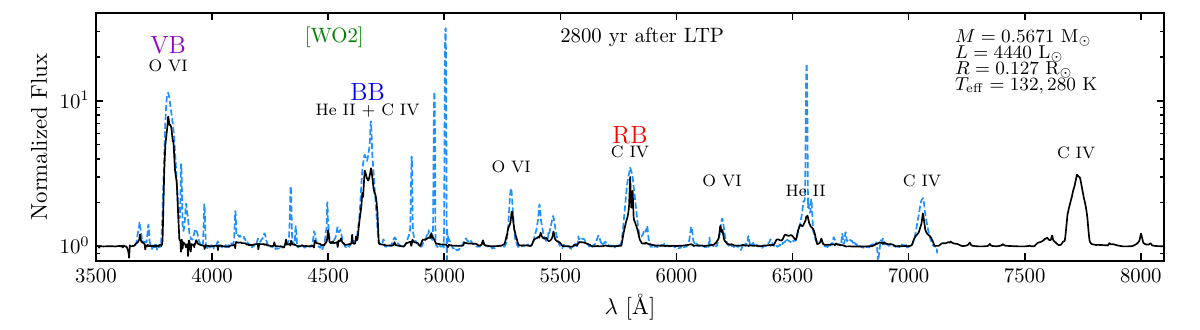}
\caption{Synthetic optical spectrum calculated with the non-LTE stellar atmosphere code {\sc PoWR} at 2800 yr of post-LTP evolution (solid line). The typical violet (VP), blue (BB), and red bump (RB) are labelled as well as other broad emission lines. The corresponding abundances are listed in the third row of Table~\ref{tab:analysis}. For comparison, the spectrum of the CSPN of NGC~6905 (dashed line), originally presented by \citet{GG2022}, is also shown. Note the contamination of the observed stellar spectrum by narrow nebular emission lines, most notably H$\beta$, H$\alpha$, and [O~{\sc iii}] $\lambda\lambda$4959,5007. }
\label{fig:spec_WO}
\end{figure*}

\section{Results}
\label{sec:results}

\subsection{The formation of an H-deficient [WR] CSPN}

After the LTP event, our stellar evolution model is able to successfully produce an H-deficient CSPN (Fig.~\ref{fig:M17_abund}). 
The model also results in similar abundances and stellar properties as those of [WR] CSPNe reported in the literature (see Fig.~\ref{fig:HRD}). For comparison, Table~\ref{tab:analysis} lists the abundance estimates for the [WR] CSPN of NGC~40, NGC~1501, NGC~2371, NGC~5189, NGC~6905, and PC~22 obtained from non-LTE stellar atmosphere modelling \citep{GG2020,GG2022,Keller2014,Rubio2022,Sabin2022,Toala2019_ngc40}. We compare the values obtained from observations with results from our model after 2235 and 2800~yr after the LTP event. That is, when the He abundance is similar to that estimated for those [WR] CSPN.
In general, the model abundances at those specific times are consistent with those measured in these [WR] stars within their uncertainties.

We note, however, that the Ne abundance of our stellar evolution model at 2235~yr is an order of magnitude larger than that
reported for [WR] CSPNe in Table~\ref{tab:analysis}.Nevertheless, by 2800 of post-LTP evolution is of the same order of magnitude as the observations.
Several works have pointed out that the Ne-rich ejecta observed in objects such as A\,30 and A\,58 are difficult to reconcile with standard born-again evolutionary models \citep{Wesson2003, Lau2011, MB2024}. In particular, \citet{MB2024} reviewed the main shortcomings of the born-again scenario and noted that the very large Ne abundances inferred for the H-deficient knots in A 30, A~58, and A~78 remain unexplained within current born-again calculations. This has led to consider alternative scenarios for the formation of born-again PNe such as a nova event on a ONeMg WD \citep{Lau2011}.  

The predicted N abundances of the model are larger than those derived for some CSPNe (e.g. NGC\,40, NGC\,2371, and PC\,22, and NGC\,6945) while lower than those derived in others (e.g. NGC 1501, Abell 30, Abell 78, Hubi 1, SwSt 1). Likewise, the N abundance is lower than that of the $M_i=1.25 M_\odot$ ($Z=0.01$) LTP model presented in \cite{MB2024}
but very close those  predicted by the $M_i=2.7 M_\odot$ ($Z=0.02$) LTP model presented in \citet{MB2005} (also shown in Table~\ref{tab:analysis}).

Fig.~\ref{fig:HRD} shows the positions of these [WR] CSPN in the HR diagram in comparison with the stellar evolution model. 
Overall, there is a notable agreement between the predicted evolutionary track and the observed properties of the sample, with most of these stars approaching or located on the turn-around point. We note, however, that our specific LTP evolution model does not exactly reproduces the luminosities of the compared [WR] CSPNe, which otherwise span over a range in $L$. 
Nevertheless, the stellar abundances associated with these sources remain in good agreement with the chemical composition predicted by our model. 

These results unambiguously confirm previous suggestions that [WR] CSPNe can be produced through a born-again event \citep{Schonberner1979,Iben1983,Herwig1999}. A situation that has been very recently found for the Sakurai's Object, where \citet{Marcolino2026} reported the emergence of a [WR]-type spectrum from this iconic born-again object. Moreover, we cannot exclude the possible role of a binary companion, which could drive the system into a common envelope phase and naturally lead to the formation of an H-deficient WR-type star \citep[e.g.,][]{Paczynski1967}. Further consideration of that mechanism, however, is beyond the scope of the present paper.

To further peer into this result, Fig.~\ref{fig:spec_WO} presents a synthetic optical spectrum obtained for the stellar and wind properties of our stellar evolution model at 2800 yr after the onset of the post-LTP phase, that is, 6,150 yr since the onset of the post-AGB phase. 
The synthetic spectrum is computed with the non-LTE stellar atmosphere {\sc PoWR} code\footnote{\url{https://www.astro.physik.uni-potsdam.de/PoWR/}} \citep{Grafener2002,Hamann2004,Todt2015b}\footnote{For a benchmark comparison of different codes used in the community see \citet{Sander2024}.}, adopting a typical clumping parameter of $D=10$ and a Fe abundance set to solar ($X_\mathrm{Fe} = 1.4\times10^{-3}$).

The synthetic spectrum exhibits the typical broad features of [WR]-type stars. 
The red bump (RB) produced mainly by the C~{\sc iv} 5806~\AA, the blue bump (BB) that includes the C~{\sc iv} 4658~\AA\, and He~{\sc ii} 4686~\AA, and the violet bump (VB) with the O~{\sc vi} 3811,3834~\AA~ doublet. 
These features are labelled on the spectrum of Fig.~\ref{fig:spec_WO} alongside other prominent broad lines.

The spectrum in Fig.~\ref{fig:spec_WO} is typical of [WO] stars, with the clear presence of O~{\sc vi} emission lines. Using the classification criteria defined by \citet{Acker2003}, this synthetic spectrum corresponds to a [WO2] spectral type.

For comparison, Fig.~\ref{fig:spec_WO} also shows the 
optical spectrum of the CSPN of NGC~6905 (dashed line) originally presented by \citet[][see specific details of the spectrum there]{GG2022}. The resemblance between model and observation is striking, once the contribution from the nebular lines is disregarded. We note that the CSPN of NGC~6905, HD 193949, has been classified as a [WO2] star.

It is important to emphasize that, although O emission lines dominate over those of C ions in [WO]-type spectra, this is purely an ionisation effect driven by their higher effective temperatures, but not the result of an O overabundance, as demonstrated in Fig.~\ref{fig:M17_abund} and Table~\ref{tab:analysis}. For further details see \citet[][]{Crowther1998} and \citet{Acker2003}.

\subsection{From a double-shell born-again PN to a [WR] PN}
\label{sec:double-shell}

\begin{figure*}
    \centering
    \includegraphics[width=1.0\linewidth]{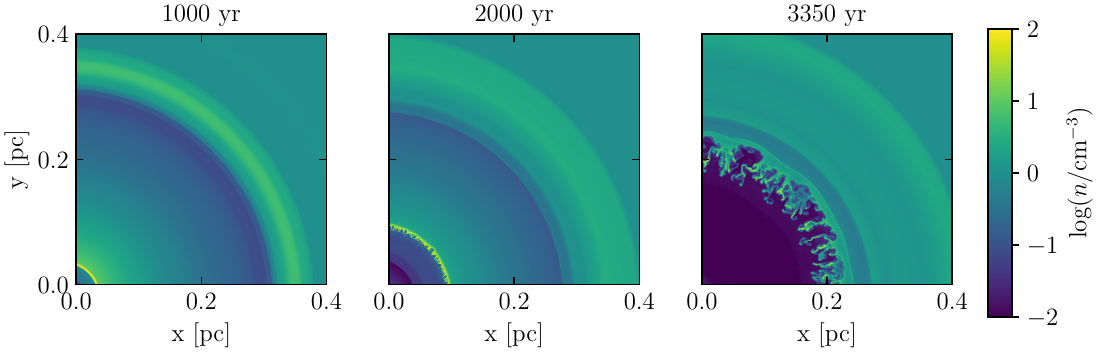}\\
    \includegraphics[width=1.0\linewidth]{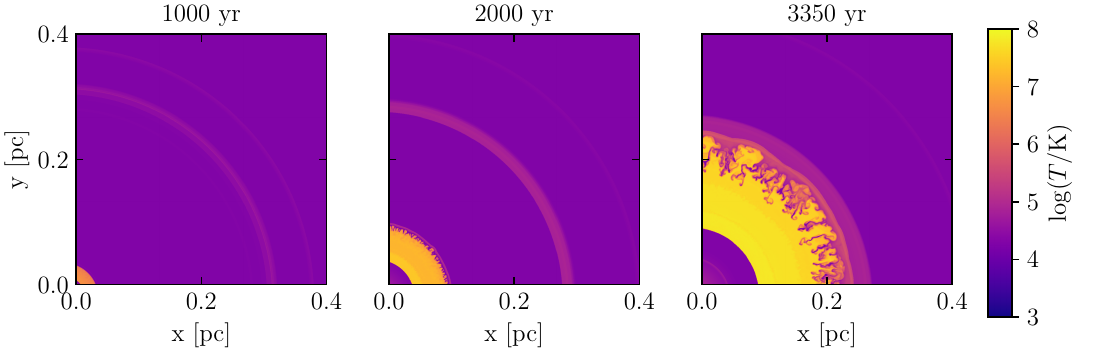}
    \caption{Evolution of the number density ($n$ - upper panel) and gas temperature ($T$ - lower panel) during the post-AGB phase. Time is measured from the beginning of the post-AGB evolution, defined as the moment when the stellar effective temperature exceeds log$_{10}(T_\mathrm{eff}$/K)=3.8.}
    \label{fig:postagb2d}
\end{figure*}

\begin{figure*}
    \centering
    \includegraphics[width=0.95\linewidth]{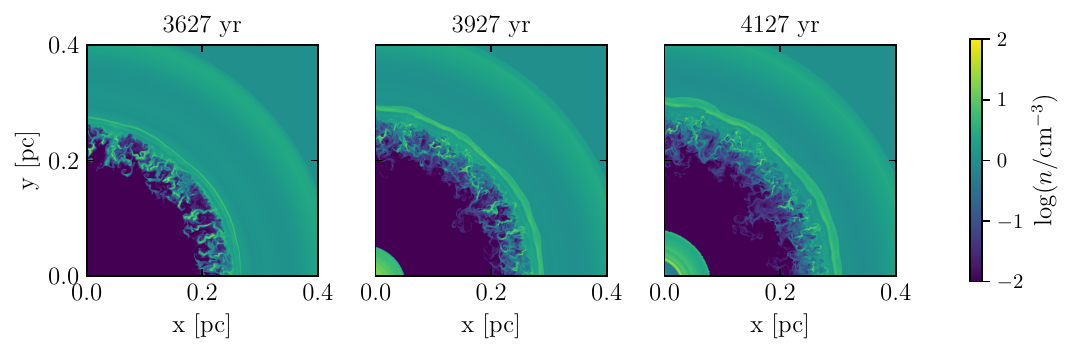}\\[-22pt]
    \includegraphics[width=0.95\linewidth]{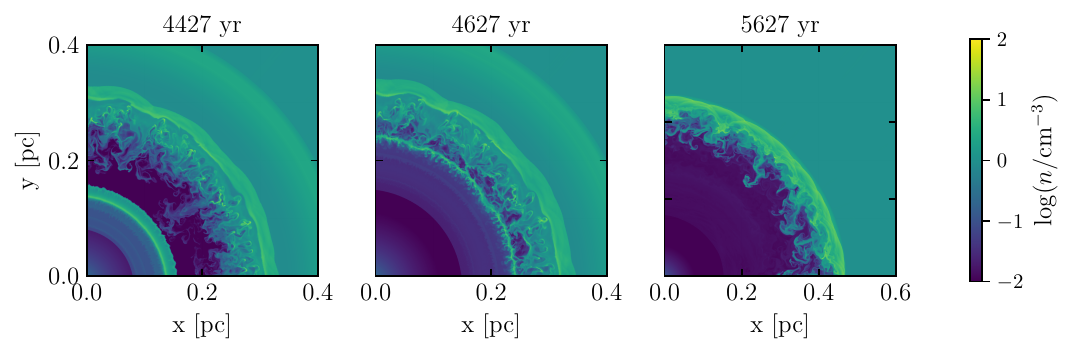}
    \caption{Temporal evolution of the number density $n$ after the LTP event. Similar to Fig.~\ref{fig:postagb2d}, $t=0$ represents the onset of the post-AGB phase. Note the different spatial scale of the bottom-right panel.}
    \label{fig:LTP-density-evolution}
\end{figure*}

\begin{figure*}
    \centering
    \includegraphics[width=0.95\linewidth]{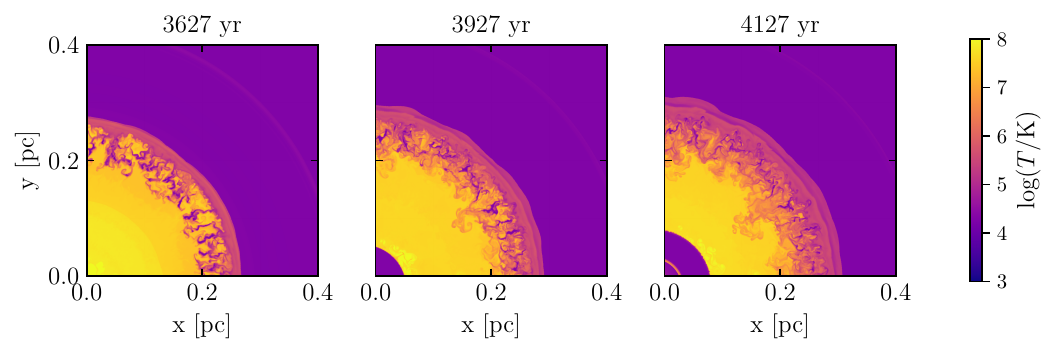}\\[-22pt]
    \includegraphics[width=0.95\linewidth]{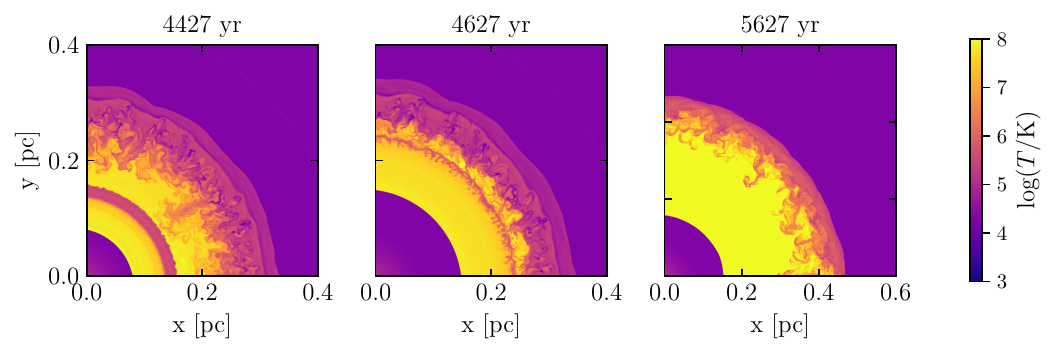}
    \caption{ Temporal evolution of the gas temperature $T$ after the LTP event. Similar to Fig.~\ref{fig:postagb2d}, $t=0$ represents the onset of the post-AGB phase. Note the different spatial scale of the bottom-right panel.}
    \label{fig:LTP-temperature-evolution}
\end{figure*}

The first stage of evolution in our simulations is that of the AGB phase, characterised by a slow wind ($\varv_\mathrm{AGB}\lesssim10$~km~s$^{-1}$) and relatively high-mass loss rate ($\dot{M}_\mathrm{AGB}\lesssim10^{-5}$~M$_\odot$~yr$^{-1}$). Such wind develops a circumstellar medium (CSM) with a density structure that follows a distribution $\sim r^{-2}$ \citep[e.g.,][]{Villaver2002a,Perinotto2004}. Details of the formation of the AGB CSM are given in Appendix~\ref{app:AGB}.

The onset of the post-AGB phase, and the beginning of the PN phase, is marked by an increase of the stellar wind velocity $\varv_\infty$, a reduction of the mass-loss rate $\dot{M}$, and the development of the ionising photon flux (see Fig.~\ref{fig:postAGB}).
The fast wind slams into the dense and slow CSM producing an adiabatic shock, a scenario that has been long studied in a variety of astrophysical systems \citep[see, e.g.,][]{Arthur2006,Villaver2002b,GarciaSegura1996,Freyer2006,Toala2011}. This produces a hot wind-blown bubble that pushes and compresses the CSM creating a dense inner rim.

\begin{figure*}
    \centering
    \includegraphics[width=0.95\linewidth]{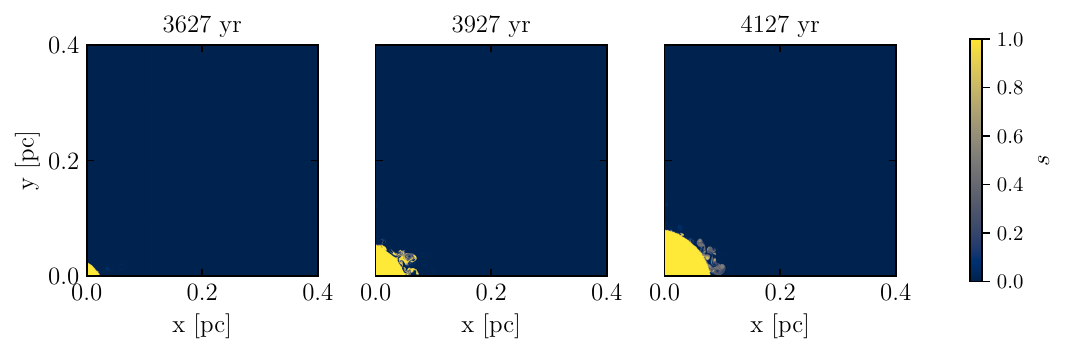}\\[-22pt]
    \includegraphics[width=0.95\linewidth]{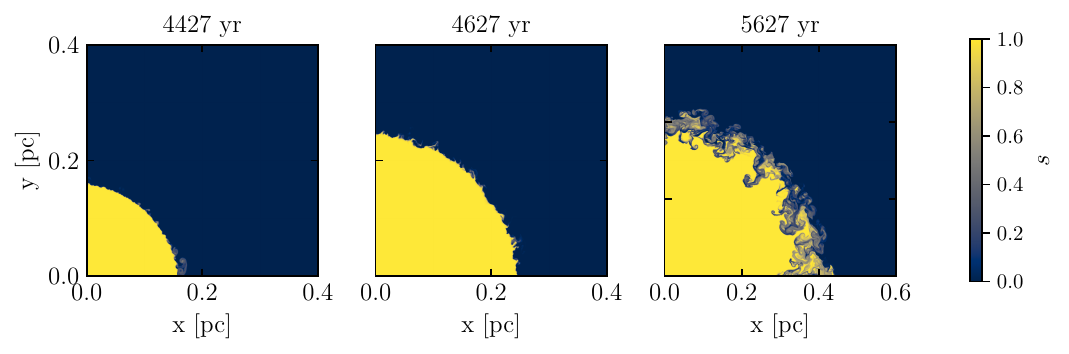}
    \caption{Temporal evolution of the advected scalar $s$ after the LTP event. Similar to Fig.~\ref{fig:postagb2d}, $t=0$ represents the onset of the post-AGB phase. Note the different spatial scale of the bottom-right panel.}
    \label{fig:scalar_plot}
\end{figure*}

The temperature of the wind-blown bubble is expected to be a function of the stellar wind velocity, $T \propto \varv_\infty^2$ \citep[see][]{Dyson1997}. However, the wind-wind interaction region is subject to hydrodynamical instabilities that naturally mix material from the ionised swept up region into the hot bubble \citep[e.g.,][]{Stute2006,Toala2014,Toala2016}, significantly changing the temperature structure of the wind-blown bubble, particularly at its outer edge\footnote{Thermal conductivity represents an additional physical process capable of producing similar effects \citep{Steffen2008,Toala2018}, but it is not accounted for in the present simulations.}. Fig.~\ref{fig:postagb2d} illustrates the evolution of the number density $n$ and gas temperature $T$ during the post-AGB phase. 

The panels show the creation of the low-density, high-temperature hot bubble and the aforementioned development of the hydrodynamical instabilities in the wind-wind interaction zone. Rayleigh-Taylor instabilities dominate the turbulent zone\footnote{Due to the existence of the instabilities, the assumption of axial symmetry is of course violated. Simulating the system in 3D would probably lead to different growth rates of clumps and filaments. However, the overall evolution of the system remains the same \citep[see, e.g.,][]{vanMarle2012}.}.

By the time star experiences the LTP event, after 3350~yr of post-AGB evolution, the fast post-AGB wind has created a hot bubble with a radius of $\gtrsim$0.2 pc (rightmost column of Fig.~\ref{fig:postagb2d}). During the LTP phase, the star reduces its mass-loss rate significantly, ejecting only $2.9\times10^{-6}$~M$_\odot$ in 277~yr (3627~yr after the onset of the post-LTP phase). Thus, there is not a significant dense shell by the end of this phase (see top left panels of Fig.~\ref{fig:LTP-density-evolution} and \ref{fig:LTP-temperature-evolution}). No significant changes are seen in the surrounding hot bubble.

The onset of the post-LTP/[WR] phase is marked by an increase in mass-loss rate, reaching values up to $10^{-6}$~M$_\odot$ for the first $\approx200$~yr of its evolution. During this time, a new fast stellar wind is developed evolving from $\sim$100 to $\sim$1000~km~s$^{-1}$ during the same period of time (see Fig.~\ref{fig:postAGB}). The combination of these effects creates a somewhat dense, H-deficient structure with 0.05~pc in radius (see middle panels of Fig.~\ref{fig:LTP-density-evolution} and \ref{fig:LTP-temperature-evolution}).
The subsequent evolution of the stellar wind is to increase the stellar wind velocity whilst decreasing the mass-loss rate. This produces an inner adiabatic hot bubble, ultimately producing a double-shell PN.
However, in just a few hundred years, the inner structure is disrupted by the action of hydrodynamical instabilities, creating an extended turbulent mixing region. The final destiny of the inner clumpy shell is to mix with the outer H-rich PN.
The evolution of the number density and temperature during the post-LTP phase is illustrated in Fig.~\ref{fig:LTP-density-evolution} and \ref{fig:LTP-temperature-evolution}, respectively.

The stellar evolution model that experiences the LTP event predicts that by $\gtrsim$1000 yr of evolution in the post-LTP phase, the H-deficient born-again ejecta will be considerably mixed with the turbulent medium at the outer edge of the first hot bubble. To illustrate the degree of mixing of the H-deficient material ejected in the LTP event with the H-rich gas from the previous stages, we included an advected passive scalar ($s$) in our simulations \citep[see][]{Toala2014,Toala2016}, which are typically used in numerical simulations to trace mixing, metallicity gradients, and fluid phases \citep[see for example][]{Colbrook2017,Drummond2020,Horst2021}. The scalar $s$ is advected alongside the velocity vector of the fluid and does not change its properties. We assigned $s=0$ to the post-AGB phase, and $s=1$ to the subsequent LTP and post-LTP phases, while intermediate values informs on the mixing of H-deficient and H-rich material. The evolution of $s$ during the post-LTP phase is shown in Fig.~\ref{fig:scalar_plot}.

The mass ejected during the LTP event is $M_\mathrm{LTP} = 2.9\times10^{-6}$ M$_\odot$, a relatively small amount compared to the total mass lost by this model 
during the AGB phase and early post-AGB phases of $\approx1.1$ M$_\odot$ \footnote{For comparison, we note that the estimates ionized mass in PNe are between 0.005 and 3 M$_\odot$ with a median value of $0.5 \times \epsilon^{1/2}$ M$_\odot$, where $\epsilon$ is a filling factor \citep{Frew2010,Frew2016}.}. Therefore, after mixing with the H-rich nebula, the LTP event should leave no significant trace. The combination of these effects is probably the explanation why PNe hosting [WR] CSPN do not exhibit significant abundance differences when compared with PNe with H-rich CSPN, both cases showing abundances consistent with models of H-rich post-AGB stars \citep[e.g.,][]{GarciaRojas2013}. Such effect hinders the identification of PNe with [WR] CSPN as part of the born-again class.

\begin{figure}
    \centering
    \includegraphics[width=1.0\linewidth]{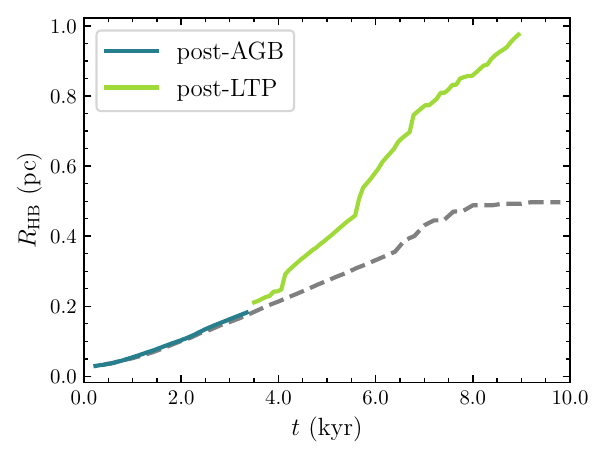}
    \caption{ Temporal evolution of the hot bubble radius ($R_\mathrm{HB}$) for a model that experiences an LTP event (solid lines). The dashed line shows the results from the nLTP model. $t=0$ yr represents the onset of the post-AGB phase.}
    \label{fig:bubble_dynamics}
\end{figure}

For comparison, the formation of a PN following the nLTP model is illustrated in Appendix~\ref{app:PN}. Fig.~\ref{fig:density_nLTP} and \ref{fig:temperature_nLTP} present the evolution of the inner hot bubble together with the development of hydrodynamical instabilities at the wind-wind interaction interface. After $\sim$8000 yr of post-AGB evolution, the nLTP model predicts a hot-bubble inner radius of approximately 0.5 pc, whereas the LTP model reaches the same radius after only $\sim$5650 yr. This difference is a direct consequence of the additional energy injected into the old PN during the LTP event. Although the stellar wind weakens substantially throughout the LTP phase, it becomes considerably stronger again during the post-LTP evolution. This behaviour is illustrated in Fig.~\ref{fig:bubble_dynamics}, which compares the temporal evolution of the hot-bubble radius ($R_\mathrm{HB}$) for both the LTP and nLTP simulations. The figure shows that, following the LTP event, the inner rim is accelerated by the enhanced post-LTP wind.

\section{Discussion}
\label{sec:discussion}

\subsection{\textbf{Stellar evolution and chemical properties}}

A major difference between our LTP model and those presented in the literature is the adoption of a stellar wind mass-loss prescription specifically tailored for the H-deficient strong born-again wind wind during the post-LTP phase. We employ the prescription of \citet{Toala2024WR}, which was originally developed to describe the winds of H-deficient [WR] stars based on fits to the wind and atmospheric properties of a sample of [WR] CSPNe and massive WR stars. In our evolutionary sequence, however, the same prescription produces a strong wind during the brief return of the star to the cool born-again AGB phase. This strong Born-Again AGB wind plays a key role in removing the remaining H-rich envelope. Once the star reheats and becomes H-deficient, the prescription naturally describes the subsequent [WR] CSPN phase.

The stronger mass-loss rates associated with the strong Born-Again wind prescription efficiently strip the remaining H-rich envelope after the LTP event, exposing deeper and hotter layers of the star at earlier times. As a consequence, the star evolves significantly faster than in the model computed with an feeble BA wind prescription (see Fig.~\ref{fig:HRD2}). While the feeble BA wind model requires several thousands of years to return to high effective temperatures, the [WR]-wind model reaches the H-deficient stage on considerably shorter timescales. The track of this model closely resembles the LTP evolutionary sequence presented by \citet{Liu2025}, especially regarding the rapid reheating phase and the morphology of the post-LTP loop in the HR diagram. The differences in the tracks and timescales further highlight the strong sensitivity of post-LTP evolution to the adopted wind prescription.

The effect of the strong Born-Again wind 
is also evident in the surface abundance evolution shown in Fig.~\ref{fig:M17_abund}. Compared to the feeble BA wind model, the strong BA wind prescription produces a much more drastic decrease in the surface H abundance shortly after the LTP event, together with a faster enrichment in He, C, and O at the stellar surface. This behaviour reflects the efficient removal of the outer H-rich layers by the intense [WR]-type mass loss.

Similarly to the LTP models presented by \citet{Liu2025} (see the lower panel of their Fig.~3), the convective regions appearing after the LTP event remain restricted to the outermost stellar layers. In contrast, before the onset of the post-LTP phase, nuclear burning regions are found deeper inside the star, while after the LTP they become confined to near-surface layers. This transition reflects the structural reconfiguration of the stellar envelope following the thermal pulse and the subsequent stripping of the outer layers by the stellar wind.

Our models exhibit substantially larger He and smaller C and O abundances during the early post-LTP evolution ($t-t_\mathrm{pLTP}\sim2800$ yr). As the post-LTP evolution proceeds, our surface abundances evolve significantly with time owing to the strong [WR]-type mass loss. The He and N abundances progressively decrease, while the C and O abundances increase, our model reaches abundances close to that presented by \citet{MB2005} and \citet{MB2024}. By 3610 yr after the onset of the post-LTP phase, our model reaches $X_\mathrm{He}\simeq0.420$, $X_\mathrm{C}\simeq0.403$, and $X_\mathrm{O}\simeq0.159$, approaching the abundance patterns predicted by those authors. They predict values in the range $X_\mathrm{He}\simeq0.31$--$0.42$, $X_\mathrm{C}\simeq0.36$--$0.43$ and $X_\mathrm{O}\simeq0.12$--$0.21$. The chemical composition of our model evolve substantially during the first 2000 years of post-LTP phase as illustrated in Fig.~\ref{fig:M17_abund}.

In contrast, the N abundances predicted by our models are larger with those reported in previous LTP studies, remaining at the level of $X_\mathrm{N}\sim10^{-3}$ throughout most of the post-LTP evolution while \citet{MB2024} reported a range of $X_\mathrm{N}\sim7.2\times10^{-5}$--$1.4\times10^{-3}$. This could be attributed to the depth of the dredge up process, that can be reflected in a slightly less O abundance as compared to the results of \citet{MB2024}.

Overall, our results point towards a post-LTP chemical composition that is systematically more He-rich and comparatively poorer in C, O, and Ne than those typically reported in previous studies. These differences highlight the strong sensitivity of the predicted surface abundances to both the adopted mass-loss prescription and the treatment of convective overshooting. In particular, our models provide a better match to the He, C, and O abundances inferred for several PNe hosting [WR] CSPNe, including NGC~40, NGC~1501, NGC~2371, NGC~5189, NGC~6905, and PC~22.

Furthermore, the relatively mild third dredge-up experienced by our models, combined with the intense mass loss following the LTP event, produces a surface composition that evolves significantly with time. Shortly after the LTP, the stellar surface is dominated by He with traces of N, reflecting the chemical composition of the He-buffer region characteristic of TP-AGB stars. At later stages, continued stripping of the outer layers exposes the intershell material, leading to the more typical He-, C-, and O-rich composition associated with [WR] CSPNe.

The duration of the LTP event has been discussed for decades and is generally expected to occur on timescales of several hundred years \citep{MB2024}. However, the exact duration predicted by stellar evolution models varies significantly depending on the adopted input physics and evolutionary assumptions. For example, \citet{Liu2025} found that the transition from the onset of the LTP to the H-deficient stage lasts approximately 800 yr in their models (see points 2 and 3 in the lower panel of their Fig.~3), whereas \citet{MB2005} obtained a considerably shorter timescale of about 100 yr. In comparison, our model predicts an intermediate duration of approximately 277 yr for the LTP event, consistent with the picture of an evolution occurring over a few centuries.

During the LTP event, the C to O abundance ratio ($X_\mathrm{C}/X_\mathrm{O}$) varies significantly, from 0.26 to values as small as 0.14 during the 277 yr of LTP evolution (see bottom panel in Fig.~\ref{fig:M17_abund}). Within 200 yr of evolution of of the post-LTP phase, the $X_\mathrm{C}/X_\mathrm{O}$ is $\gtrsim 1$. This behaviour is expected from the dredge-up and envelope enrichment that follow late thermal pulses, and is consistent with previous born-again and post-AGB evolutionary calculations, which predict C-rich surface compositions and $X_\mathrm{C}/X_\mathrm{O} > 1$ after a final He-shell flash \citep[e.g.][]{Iben1983,Herwig2001,Lawlor2003,MB2006}. During that time, the star is located in the $T_\mathrm{eff} < 10^{4}$ K region of the HR diagram, allowing the formation of C-rich dust in the born-again ejecta. The predicted $X_\mathrm{C}/X_\mathrm{O}$ is consistent with the value of 1.27 obtained by photoionisation modelling including C-rich dust and gas presented in \citet{Toala2021dust} for the born-again PN A~30. 
Nevertheless, it should be noted that our model may not reproduce all the observed properties of A~30 and A~78, since both have been suggested to be descendants of a VLTP event, which is not the born-again event modelled here.

Nevertheless, it should be noted that our model may not reproduce all the observed properties of A~30 and A~78, since both have been suggested to be descendants of a VLTP event, which is not the born-again event modelled here.

Among the [WR] CSPNe listed in Table~\ref{tab:analysis}, that of NGC~5189 stands out because of its larger $X_\mathrm{N}$. This CSPN has a larger nitrogen abundance ($X_\mathrm{N} = 0.01$) compared with the values reported for other [WR] CSPN ($X_\mathrm{N} \leq 7\times10^{-3}$), which are more consistent with the predictions of LTP models. 
According to the summary of born-again models gathered in \citet{MB2024}, the nitrogen abundance of NGC~5189 seems to be more consistent with a VLTP event. Such differences seem to suggest that $X_\mathrm{N}$ can be used to discern between an LTP and a VLTP event.

\begin{figure*}
\begin{center}
\includegraphics[width=\linewidth]{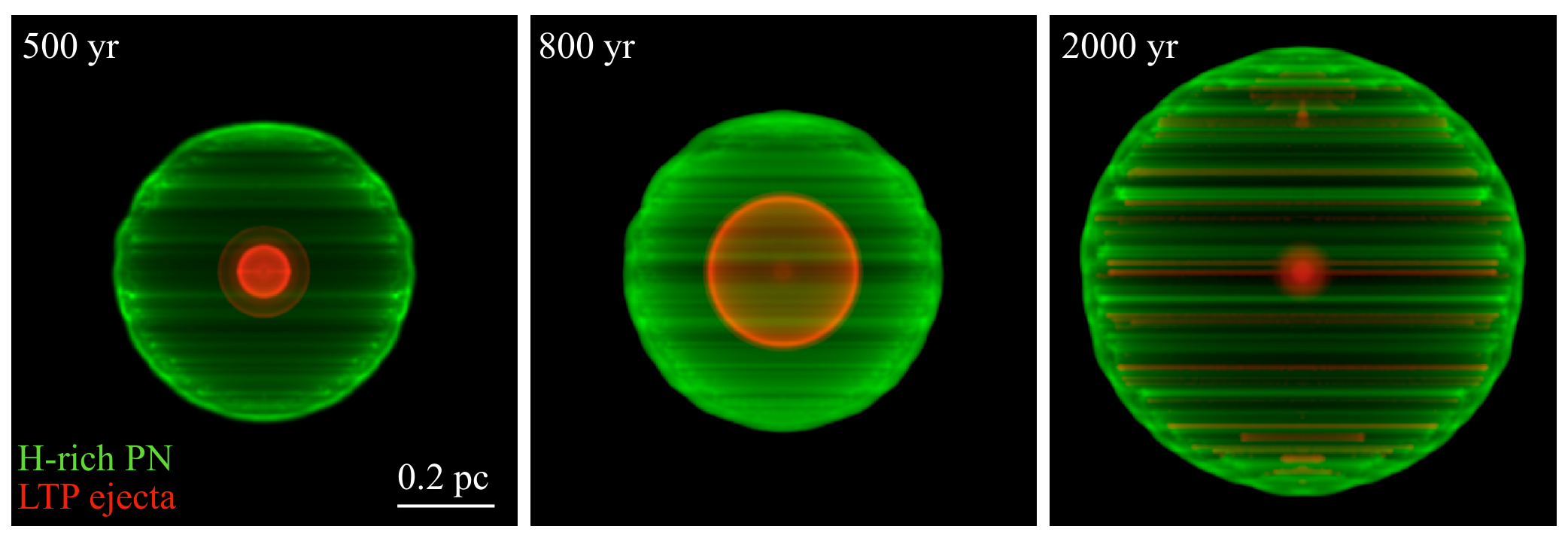}
\end{center}
\caption{Synthetic nebular images at 500, 800 and 2000 yr after the LTP event. The green and red colours correspond to H-rich nebular material ($s < 0.5$) and H-deficient LTP ejecta ($s \geq 0.5$), respectively. The images were produced convolving the density distribution with the scalar information.}
\label{fig:synt_neb}
\end{figure*}

\subsection{Observables of born-again PNe}

The simulations presented in Section~\ref{sec:double-shell} confirmed that the natural evolution of a star experiencing an LTP is the formation of a double-shell PNe. With the inner shell having H-deficient abundances. 
Such H-deficient structures have been detected in well-known born-again PNe A~30, A~58, A~78, and HuBi~1 \citep[see][and references therein]{Clayton2013,Fang2014,MontoroMolina2022}.
It is straightforward to classify a PN as part of the born-again family if it hosts a [WR]-type CSPN and exhibits a double-shell structure, with the inner shell being H-deficient. However, our simulations predict that the double-shell morphology of born-again PNe should be short-lived. The present LTP model predicts $\gtrsim 1000$ yr for the survival of the H-deficient shell.

To further illustrate the evolution of the H-deficient LTP ejecta and its interaction with the old PN, we created synthetic nebular images using the density and scalar distributions from the simulations. The images of the H-rich nebula were created by combining the density maps with scalar values $s < 0.5$, while the LTP ejecta images were created by accounting for scalar values $s \geq 0.5$. Three snapshots at 500, 800, and 2000 yr after the LTP event are shown in Fig.~\ref{fig:synt_neb}. 
The figure shows that at early times the H-deficient ejecta has a clear shell-like structure with extended emission, resembling the morphology seen in HuBi~1 \citep{Guerrero2018}. As the LTP ejecta expands its surface brightness is reduced (see middle panel). Finally, after 2000~yr of evolution the disrupted H-poor nebula has been disrupted and mixed with the outer H-rich structure. 

The rightmost panel in the figure is reminiscent of NGC~1501 \citep{Rubio2022}
\footnote{It is important to remark here that the nebular images were created from 2D simulations. In order to estimate the emissivity of 3D structures we rotated the 2D simulations over their own symmetry axis. This produces the numerical artifacts shown as horizontal straight structures.}.

It is worth mentioning here that the H-deficient material in known born-again PNe has a clear bipolar distribution. In most cases, the H-deficient ejecta is distributed in a (disrupted) disc plus a pair of jet-like features. These structures have been detected in optical using {\it Hubble Space Telescope} data for A~30 and A~78 \citep{Borkowski1993,Borkowski1995,Fang2014}, in sub-millimetre data from ALMA in the case of A~58 and the Sakurai's Object \citet{Tafoya2022,Tafoya2023}, and have been suggested in the analysis of IR data in the case of NGC~40 \citep{Toala2019_ngc40}. Thus, we are yet to confirm if the born-again event is not isotropically ejected from the surface of the star or if a companion helps shaping the H-deficient material into bipolar structures \citep[e.g.,][]{Soker1997,RodriguezGonzalez2022}. 
Such mechanisms are not explored here, because given their increase in complexity will require the exploration of 3D radiation-hydrodynamical simulations with non-spherical inner boundary conditions and, in the binary case, an explicit treatment of orbital interaction and angular momentum transfer.

\subsubsection{Turbulence and dynamical impact of the LTP event}

To further assess the dynamical impact of the LTP event, we analyse the evolution of the turbulent kinetic energy (TKE) in the simulations. The TKE represents the energy contained in chaotic non-radial gas motions and therefore provides a quantitative measure of the strength of hydrodynamical instabilities developing at the wind-shell interface.

To compute the TKE, we perform a Reynolds decomposition of the velocity field into a mean radial component and a fluctuating component,
\begin{equation}
    \mathbf{v} = \langle \mathbf{v} \rangle + \mathbf{v}',
\end{equation}
where $\langle \mathbf{v} \rangle$ corresponds to the mean radial flow and $\mathbf{v}'$ represents the velocity fluctuations associated with turbulent motions. The turbulent kinetic energy density is then defined as
\begin{equation}
    \epsilon_{\rm turb} = \frac{1}{2}\rho |\mathbf{v}'|^{2},
\end{equation}
where $\rho$ is the local gas volume density. The total turbulent kinetic energy is obtained by integrating this quantity over the simulation volume,
\begin{equation}
    E_{\rm turb} = \int \epsilon_{\rm turb}\, dV,
\end{equation}
accounting for the axisymmetric geometry of the simulations.

\begin{figure}
    \centering
    \includegraphics[width=1.0\linewidth]{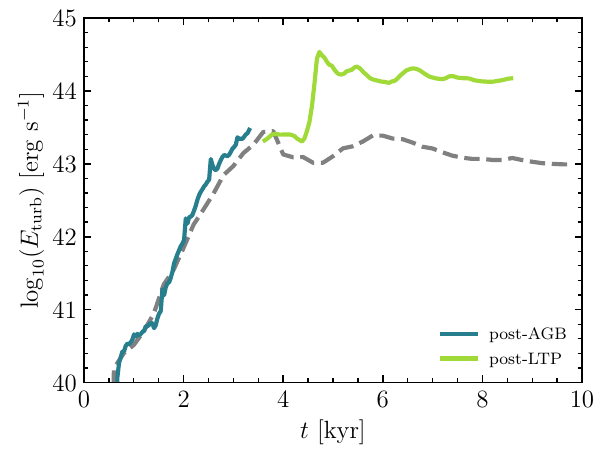}
    \caption{Total turbulent kinetic energy (TKE) in units of erg. The $1.7~M_{\odot}$ model is separated into the initial post-AGB phase (blue solid line) and the subsequent Post-LTP phase (green solid line). The results from the nLTP model are shown with a dashed line.}
    \label{fig:turbulent_energy}
\end{figure}

Fig.~\ref{fig:turbulent_energy} shows the temporal evolution of the total TKE for our LTP model (solid lines) and that resulted from the nLTP simulation (dashed line). Similarly to that illustrated in Fig.~\ref{fig:bubble_dynamics}, during the early post-AGB phase, both models exhibit a similar behaviour with an increase in turbulent energy, indicating that hydrodynamical instabilities develop in a comparable manner in the absence of an LTP.

By the time the star experiences the LTP, the two models diverge dramatically. While the standard PN model maintains a nearly constant turbulent energy of $\sim$ 10$^{43}$\,erg, the post-LTP evolution results in an order of magnitude larger TKE.

In addition to Fig.~\ref{fig:bubble_dynamics}, the analysis presented in this section confirms that the high mechanical luminosity of the post-LTP strong Born-Again wind produces a highly turbulent environment and a faster expansion of the inner rim. Characteristics that have been reported from the analysis of observations of [WR] PNe when compared with PNe with H-rich CSPN \citep{Gesicki2006,Medina2006}.

This high degree of turbulent mixing may also help explain why [WR] PNe are more frequently detected by X-ray instruments \citep{Freeman2014}. In hydrodynamical models of PNe, the observed soft X-ray emission is understood to arise from gas at intermediate temperatures produced at the interface between the hot shocked wind and the cooler nebular material, where mixing and/or thermal conduction play a key role \citep{Schonberner2024,Orozco-Duarte2026}.

\subsubsection{Contribution from LTP gas and observability of the born-again phenomenon}

While the hydrodynamical evolution described above demonstrates that the H-deficient ejecta is rapidly mixed with the pre-existing surrounding nebular gas, an important question remains: to what extent does this material remain observable once the mixing has taken place? In other words, the mere presence of the H-deficient gas does not necessarily implies that its signature can be detected in the emitted radiation.

In photoionised nebulae, the observable emission is weighted towards the densest regions, since it scales approximately with the square of the gas density. As a consequence, the emission arising from the LTP ejecta, as it expands into the pre-existing cavity carved by the stellar winds of the post-AGB star, is expected to contribute significantly to the overall emission of the born-again PNe.

To quantify this effect, we make use of the advected scalar $s$, which traces the fraction of gas originating from the H-poor LTP ejecta. By combining this tracer with the local gas density, we define an emission-weighted diagnostic that measures the relative contribution of the H-deficient material to the total nebular emission. In this framework, we define the emission-weighted fraction of LTP material as:
\begin{equation}
    f_{\rm LTP}(t) = 
    \frac{\int s(\mathbf{x},t)\, n^2(\mathbf{x},t)\, dV}
         {\int n^2(\mathbf{x},t)\, dV},
\end{equation}
\noindent where $n$ is the gas number density and $dV$ is the volume element associated with each computational cell. This quantity provides a proxy for the contribution of H-deficient material to the observable nebular emission, under the assumption that the emissivity scales with the square of the gas density. Although this diagnostic does not correspond to a full line-of-sight projection, it provides a useful approximation of the observability of the H-deficient material in an axisymmetric framework.

\begin{figure}
    \centering
    \includegraphics[width=1.0\linewidth]{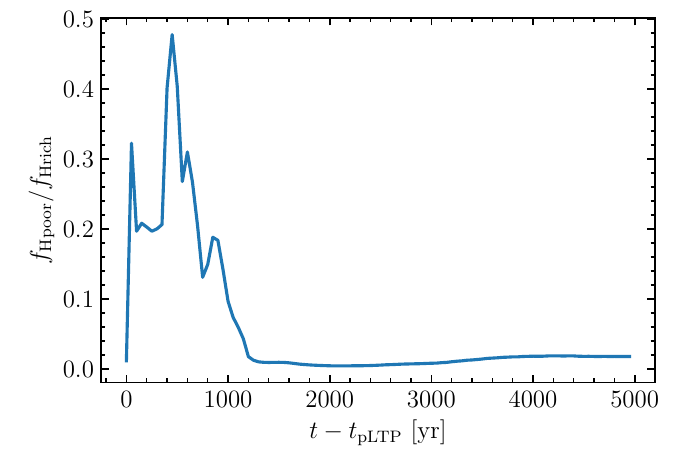}
    \caption{Temporal evolution of the emission-weighted fraction of LTP material, $f_{\rm LTP}$, as defined by the density-squared weighting of the advected scalar tracing the H-deficient ejecta. The time is measured from the onset of the post-LTP phase. At early times, the LTP ejecta contributes significantly to the total emission, reaching a maximum due to compression by the fast post-LTP wind. As hydrodynamical instabilities develop, the ejecta is progressively fragmented and mixed with the surrounding nebular gas, leading to a rapid decline in $f_{\rm LTP}$. Within $\sim 10^{3}$~yr, the emission-weighted contribution of the H-deficient material becomes negligible.}
    \label{fig:emission_fraction_LTP}
\end{figure}

Fig.~\ref{fig:emission_fraction_LTP} shows the emission-weighted fraction of LTP material as a function of time. The $x$-axis represents the simulation time measured from the onset of the post-LTP phase. At times earlier than 1000~yr, the value of $f_\mathrm{Hpoor}$ is $\approx0.3$, reaching a maximum at $\sim400$~yr, indicating that the emission of the H-deficient material is significant. The maximum value corresponds to the phase in which the H-deficient ejecta is further compressed by the increase in velocity of the fast post-LTP stellar wind. This interaction leads to the formation of dense structures, which dominate the emission due to the $n^{2}$ weighting. As a result, the LTP material temporarily provides a significant contribution to the total nebular emission.

However, this contribution is short-lived. As hydrodynamical instabilities develop, the ejecta becomes fragmented into clumps and filaments that are progressively ablated and mixed with the surrounding H-rich nebular gas. Consequently, the emission-weighted contribution of the LTP material decreases rapidly, dropping below 10\% within $\sim 1000$ yr of evolution in the post-LTP phase.

This behaviour is consistent with the dynamical evolution described above: an initial phase characterised by the formation of a dense inner shell, followed by its compression by the fast wind, the development of instabilities, and the subsequent disruption and mixing of the H-deficient material. As the mixing proceeds, the H-rich nebular gas increasingly dominates the emission, effectively erasing the observable signature of the born-again event.

\subsection{Comparisons with previous numerical works}

The number of numerical studies addressing the formation and evolution of born-again PNe remains relatively limited. Moreover, none of the previous works followed the detailed stellar evolution of a star undergoing a born-again event in a self-consistent manner, as done in the present study.

Early radiation-hydrodynamical simulations by \citet{Fang2014} focused on reproducing the morphology and expansion patterns of the H-poor knots observed in evolved born-again PNe such as A~30 and A~78. Their results showed that hydrodynamical instabilities naturally fragment the ejecta into clumps and filaments, highlighting the complex interaction between the fast stellar wind and the H-deficient material, including mass loading and ablation processes. Those simulations exhibit the fragmentation at earlier times than in the results presented here, as a consequence of the faster increase of the wind velocity in the post-born-again evolution of the simulations presented in \citet{Fang2014}.
Moreover, in that work the evolution of the CSPN was prescribed to approximately mimic the expected evolution of a VLTP event, rather than being derived from detailed stellar evolution calculations.

More recently, \citet{Toala2021} presented 3D radiation-hydrodynamical simulations aimed at reproducing the properties of the born-again PN HuBi~1. Their study demonstrated that the interaction between multiple mass-ejection episodes can naturally generate structures similar to those observed in that nebula. Owing to the uncertainties in the wind velocity during the born-again phase, \citet{Toala2021} explored two different wind configurations and concluded that, although the resulting kinematic signatures differ, the overall morphology remains broadly similar. In addition, we note that the numerical resolution was not enough to assess the formation of hydrodynamical instabilities.

Radiation-hydrodynamical models by \citet{Ary2023} also investigated the effects of multiple mass-ejection episodes on the ionization and dynamical structure of HuBi~1. Although their code includes an improved treatment of atomic and ionic processes, the 1D nature of their simulations prevents the development of hydrodynamical instabilities and turbulent structures enhanced during the born-again event.

In this context, our work complements previous studies by providing multidimensional radiation-hydrodynamical simulations of the LTP scenario based on detailed stellar evolution calculations. In particular, we explicitly quantify the mixing efficiency between the H-deficient ejecta and the pre-existing nebular material, as well as its impact on the observability of born-again signatures. Our results suggest that the interaction between the fast post-LTP wind and the dense ejecta drives enhanced turbulence and rapid mixing, with characteristic timescales of $\sim 10^{3}$ yr, significantly affecting the long-term detectability of H-deficient material in born-again PNe.

\section{Summary and concluding remarks}
\label{sec:summary}

We presented the first hydrodynamical numerical simulations that addressed the formation of born-again PNe through an LTP, following in detail the results from a stellar evolution model produced with {\sc mesa}. 

The stellar evolution model presented here corresponds to a $M_\mathrm{ZAMS} = 1.7$ M$_\odot$ and $Z_\mathrm{ZAMS} = 0.02$ without rotation or magnetic field effects. The model incorporated {\bf a }key improvement over previous works{\bf ,} we adopted a mass-loss prescription during the post-LTP phase tailored for H-deficient [WR]-type winds. 
This stellar evolution model is then used as input for stellar atmosphere non-LTE models with {\sc PoWR} and axially symmetric 2D radiation-hydrodynamical simulations performed with the extensively tested code {\sc pluto}.

Our main conclusions can be summarized as follows:

\begin{itemize}

\item Our stellar evolution model predicts abundances slightly different from previous LTP models. However, these new predictions are very similar to those of [WR] CSPN reported in the literature. In particular, a virtually zero H abundance is reached as the result of the strong mass-loss rate during the post-LTP phase. Overall, our model reproduces well the observed abundances of H-deficient [WR] CSPN. Our predictions compare very favourably with the properties of the [WR] CSPNe of NGC~40, 
NGC~1501, NGC~2371, NGC~5189, NGC~6905, and PC~22.

\item A synthetic optical spectrum computed 2800 yr after the LTP event predicts that after the born-again event the H-deficient CSPN will evolve extremely fast into a [WO]-type star. This strengthen the view that born-again events are a viable pathway for producing H-deficient [WR] stars.

\item The stellar evolution model predicts a carbon-to-oxygen abundance ratio $X_\mathrm{C} / X_\mathrm{O} \gtrsim 1.0$ at late stages of the LTP evolution. In conjunction with the star evolution to low $T_\mathrm{eff}$, it can be expected the formation of large amounts of C-rich dust during this short phase. This behaviour is similar to that seen in PNe that have experienced a VLTP event, as shown by photoionisation modelling including both C-rich dust and gas of A~30. 
\item Our radiation-hydrodynamical simulations show that a star forming a PN and subsequently undergoing an LTP event develops a short-lived double-shell structure. Our model indicates that, after roughly 1000~yr of post-LTP evolution, the H-deficient ejecta will mix with the clumps and filaments formed in the innermost nebular rim. 
The total mass of the LTP ejecta is 2.9$\times$10$^{-6}$~M$_\odot$, which is relatively small compared to the mass of the ionised H-rich nebula. Consequently, once mixing occurs, it becomes impossible to disentangle or identify the presence of the H-deficient ejecta. 
This provides an explanation for PNe hosting [WR] CSPNe with nebular abundances that agree with models of H-rich post-AGB stars.

\item The LTP event imprints kinetic energy into the old H-rich causing a faster expansion compared with models that do not experience a born-again event. In addition, the interaction of the H-deficient ejecta with the inner rim of the H-rich nebula produces a layer of enhanced turbulence. 
These results are in line with previous observational evidence that [WR] PNe exhibit larger expansion velocities and turbulence than PNe hosting H-rich CSPN.

\end{itemize}

\section*{Acknowledgements} 
 
The authors thank the anonymous referee 
for comments and suggestions that improved the presentation of our results. J.B.R.G., R.O.D., and J.A.T.~acknowledge support from the UNAM PAPIIT project IN102324. J.A.T.~thanks the staff of Facultad de Ciencias de la Tierra y el Espacio of Universidad Aut\'{o}noma de Sinaloa (FACITE-UAS, Mexico) for their support during a research visit. 
J.B.R.G.\ and M.A.G.\ acknowledge financial support from grants CEX2021-001131-S funded by MCIN/AEI/10.13039/501100011033 and PID2022-142925NB-I00 from the Spanish Ministerio de Ciencia, Innovaci\'on y Universidades (MCIU) co-funded with FEDER funds. J.B.R.G.\ and M.A.G.\ also  acknowledge financial support from the Severo Ochoa grant CEX2021-001131-S funded by MCIN/AEI/ 10.13039/501100011033.
M3B acknowledges support by the CONICET-DAAD 2022 bilateral cooperation grant number 80726, and grants PIP-2971 and PICT 2020-03316 by  CONICET and Agencia I+D+i. 
R.K.~acknowledges financial support via the Heisenberg Research Grant funded by the Deutsche Forschungsgemeinschaft (DFG, German Research Foundation) under grant no.~KU 2849/9, project no.~445783058. 
This work has made extensive use of NASA's Astrophysics Data System.

%%%%%%%%%%%%%%%%%%%%%%%%%%%%%%%%%%%%%%%%%%%%%%%%%%



%\bibliographystyle{mnras}

\section*{DATA AVAILABILITY}

The MESA stellar evolution model, including the input files required to reproduce the calculations presented in this work, is publicly available through Zenodo at DOI: \url{10.5281/zenodo.20719170
}. The hydrodynamical calculations presented in this article will be shared on reasonable request to the corresponding author. %\textcolor{red}{Don't forget to incliude Zenodo link.}

\bibliographystyle{mnras}
\bibliography{references}

@BOOK{2013KippBook,
       author = {{Kippenhahn}, Rudolf and {Weigert}, Alfred and {Weiss}, Achim},
        title = "{Stellar Structure and Evolution}",
         year = 2013,
          doi = {10.1007/978-3-642-30304-3},
       adsurl = {https://ui.adsabs.harvard.edu/abs/2013sse..book.....K}
}

@ARTICLE{Marcolino2026,
       author = {{Marcolino}, W. and {van Hoof}, P.~A.~M. and {Maria}, V.~C. and {Todt}, H. and {Van de Steene}, G.~C. and {Toal{\'a}}, J.~A. and {Kimeswenger}, S. and {Hajduk}, M. and {Bouret}, J.-C. and {Tafoya}, D. and et al.},
        title = "{The emergence of a [WC] star in Sakurai's object}",
      journal = {\mnras},
         year = 2026,
        month = oct,
       volume = {552},
       number = {1},
          eid = {stag1533},
        pages = {stag1533},
          doi = {10.1093/mnras/stag1533},
archivePrefix = {arXiv},
       eprint = {2609.18645},
 primaryClass = {astro-ph.SR},
       adsurl = {https://ui.adsabs.harvard.edu/abs/2026MNRAS.552g1533M}
}

@ARTICLE{GarciaSegura1997,
       author = {{Garcia-Segura}, Guillermo},
        title = "{Three-dimensional Magnetohydrodynamical Modeling of Planetary Nebulae: The Formation of Jets, Ansae, and Point-symmetric Nebulae via Magnetic Collimation}",
      journal = {\apjl},
         year = 1997,
        month = nov,
       volume = {489},
        pages = {L189},
          doi = {10.1086/316796},
       adsurl = {https://ui.adsabs.harvard.edu/abs/1997ApJ...489L.189G}
}

@ARTICLE{Frank1994,
       author = {{Frank}, Adam},
        title = "{The Unity and Diversity of Planetary or Nebulae Radiation- Gasdynamics of PNe. II.}",
      journal = {\aj},
         year = 1994,
        month = jan,
       volume = {107},
        pages = {261},
          doi = {10.1086/116850},
       adsurl = {https://ui.adsabs.harvard.edu/abs/1994AJ....107..261F}
}

@ARTICLE{GarciaSegura1996,
       author = {{Garcia-Segura}, G. and {Langer}, N. and {Mac Low}, M.-M.},
        title = "{The hydrodynamic evolution of circumstellar gas around massive stars. II. The impact of the time sequence O star -> RSG -> WR star.}",
      journal = {\aap},
         year = 1996,
        month = dec,
       volume = {316},
        pages = {133-146},
       adsurl = {https://ui.adsabs.harvard.edu/abs/1996A&A...316..133G}
}

@ARTICLE{Freyer2006,
       author = {{Freyer}, Tim and {Hensler}, Gerhard and {Yorke}, Harold W.},
        title = "{Massive Stars and the Energy Balance of the Interstellar Medium. II. The 35 M$_{solar}$ Star and a Solution to the ``Missing Wind Problem''}",
      journal = {\apj},
         year = 2006,
        month = feb,
       volume = {638},
       number = {1},
        pages = {262-280},
          doi = {10.1086/498734},
archivePrefix = {arXiv},
       eprint = {astro-ph/0512110},
 primaryClass = {astro-ph},
       adsurl = {https://ui.adsabs.harvard.edu/abs/2006ApJ...638..262F}
}

@ARTICLE{Frank1993,
       author = {{Frank}, Adam and {Balick}, Bruce and {Icke}, Vincent and {Mellema}, Garrelt},
        title = "{Astrophysical Gasdynamics Confronts Reality: The Shaping of Planetary Nebuale}",
      journal = {\apjl},
         year = 1993,
        month = feb,
       volume = {404},
        pages = {L25},
          doi = {10.1086/186735},
       adsurl = {https://ui.adsabs.harvard.edu/abs/1993ApJ...404L..25F}
}

@ARTICLE{Mellema1994,
       author = {{Mellema}, G.},
        title = "{The gasdynamic evolution of spherical planetary nebulae. Radiation-gasdynamics of PNe III.}",
      journal = {\aap},
         year = 1994,
        month = oct,
       volume = {290},
        pages = {915-935},
          doi = {10.48550/arXiv.astro-ph/9405048},
archivePrefix = {arXiv},
       eprint = {astro-ph/9405048},
 primaryClass = {astro-ph},
       adsurl = {https://ui.adsabs.harvard.edu/abs/1994A&A...290..915M}
}

@ARTICLE{Mellema1997,
       author = {{Mellema}, G.},
        title = "{The formation of bipolar planetary nebulae.}",
      journal = {\aap},
         year = 1997,
        month = may,
       volume = {321},
        pages = {L29-L32},
          doi = {10.48550/arXiv.astro-ph/9704172},
archivePrefix = {arXiv},
       eprint = {astro-ph/9704172},
 primaryClass = {astro-ph},
       adsurl = {https://ui.adsabs.harvard.edu/abs/1997A&A...321L..29M}
}

@ARTICLE{GarciaSegura2025,
       author = {{Garc{\'\i}a-Segura}, Guillermo and {Manchado}, Arturo and {Toal{\'a}}, Jes{\'u}s A. and {Guerrero}, Mart{\'\i}n A. and {Castro-Tirado}, Alberto J.},
        title = "{Planetary nebula evolution for single stellar models. The formation of neutral spikes}",
      journal = {\mnras},
         year = 2025,
        month = nov,
       volume = {543},
       number = {4},
        pages = {3867-3884},
          doi = {10.1093/mnras/staf1744},
archivePrefix = {arXiv},
       eprint = {2510.08721},
 primaryClass = {astro-ph.SR},
       adsurl = {https://ui.adsabs.harvard.edu/abs/2025MNRAS.543.3867G}
}

@ARTICLE{Toala2021,
       author = {{Toal{\'a}}, J.~A. and {Lora}, V. and {Montoro-Molina}, B. and {Guerrero}, M.~A. and {Esquivel}, A.},
        title = "{Formation and fate of the born-again planetary nebula HuBi 1}",
      journal = {\mnras},
         year = 2021,
        month = aug,
       volume = {505},
       number = {3},
        pages = {3883-3891},
          doi = {10.1093/mnras/stab1592},
archivePrefix = {arXiv},
       eprint = {2103.11503},
 primaryClass = {astro-ph.SR},
       adsurl = {https://ui.adsabs.harvard.edu/abs/2021MNRAS.505.3883T}
}

@ARTICLE{Drummond2020,
       author = {{Drummond}, Benjamin and {H{\'e}brard}, Eric and {Mayne}, Nathan J. and {Venot}, Olivia and {Ridgway}, Robert J. and {Changeat}, Quentin and {Tsai}, Shang-Min and {Manners}, James and {Tremblin}, Pascal and {Abraham}, Nathan Luke and {Sing}, David and {Kohary}, Krisztian},
        title = "{Implications of three-dimensional chemical transport in hot Jupiter atmospheres: Results from a consistently coupled chemistry-radiation-hydrodynamics model}",
      journal = {\aap},
         year = 2020,
        month = apr,
       volume = {636},
          eid = {A68},
        pages = {A68},
          doi = {10.1051/0004-6361/201937153},
archivePrefix = {arXiv},
       eprint = {2001.11444},
 primaryClass = {astro-ph.EP},
       adsurl = {https://ui.adsabs.harvard.edu/abs/2020A&A...636A..68D}
}

@ARTICLE{Colbrook2017,
       author = {{Colbrook}, Matthew J. and {Ma}, Xiangcheng and {Hopkins}, Philip F. and {Squire}, Jonathan},
        title = "{Scaling laws of passive-scalar diffusion in the interstellar medium}",
      journal = {\mnras},
         year = 2017,
        month = may,
       volume = {467},
       number = {2},
        pages = {2421-2429},
          doi = {10.1093/mnras/stx261},
archivePrefix = {arXiv},
       eprint = {1610.06590},
 primaryClass = {astro-ph.GA},
       adsurl = {https://ui.adsabs.harvard.edu/abs/2017MNRAS.467.2421C}
}

@ARTICLE{Horst2021,
       author = {{Horst}, L. and {Hirschi}, R. and {Edelmann}, P.~V.~F. and {Andr{\'a}ssy}, R. and {R{\"o}pke}, F.~K.},
        title = "{Multidimensional low-Mach number time-implicit hydrodynamic simulations of convective helium shell burning in a massive star}",
      journal = {\aap},
         year = 2021,
        month = sep,
       volume = {653},
          eid = {A55},
        pages = {A55},
          doi = {10.1051/0004-6361/202140825},
archivePrefix = {arXiv},
       eprint = {2107.02199},
 primaryClass = {astro-ph.SR},
       adsurl = {https://ui.adsabs.harvard.edu/abs/2021A&A...653A..55H}
}

@ARTICLE{Toala2021dust,
       author = {{Toal{\'a}}, J.~A. and {Jim{\'e}nez-Hern{\'a}ndez}, P. and {Rodr{\'\i}guez-Gonz{\'a}lez}, J.~B. and {Estrada-Dorado}, S. and {Guerrero}, M.~A. and {G{\'o}mez-Gonz{\'a}lez}, V.~M.~A. and {Ramos-Larios}, G. and {Garc{\'\i}a-Hern{\'a}ndez}, D.~A. and {Todt}, H.},
        title = "{Carbon dust in the evolved born-again planetary nebulae A 30 and A 78}",
      journal = {\mnras},
         year = 2021,
        month = may,
       volume = {503},
       number = {1},
        pages = {1543-1556},
          doi = {10.1093/mnras/stab593},
archivePrefix = {arXiv},
       eprint = {2102.12884},
 primaryClass = {astro-ph.GA},
       adsurl = {https://ui.adsabs.harvard.edu/abs/2021MNRAS.503.1543T}
}

@ARTICLE{Hamann2004,
       author = {{Hamann}, W. -R. and {Gr{\"a}fener}, G.},
        title = "{Grids of model spectra for WN stars, ready for use}",
      journal = {\aap},
         year = 2004,
        month = nov,
       volume = {427},
        pages = {697-704},
          doi = {10.1051/0004-6361:20040506},
       adsurl = {https://ui.adsabs.harvard.edu/abs/2004A&A...427..697H}
}

@ARTICLE{Todt2015b,
       author = {{Todt}, H. and {Sander}, A. and {Hainich}, R. and {Hamann}, W. -R. and {Quade}, M. and {Shenar}, T.},
        title = "{Potsdam Wolf-Rayet model atmosphere grids for WN stars}",
      journal = {\aap},
         year = 2015,
        month = jul,
       volume = {579},
          eid = {A75},
        pages = {A75},
          doi = {10.1051/0004-6361/201526253},
       adsurl = {https://ui.adsabs.harvard.edu/abs/2015A&A...579A..75T}
}

@ARTICLE{Grafener2002,
       author = {{Gr{\"a}fener}, G. and {Koesterke}, L. and {Hamann}, W. -R.},
        title = "{Line-blanketed model atmospheres for WR stars}",
      journal = {\aap},
         year = 2002,
        month = may,
       volume = {387},
        pages = {244-257},
          doi = {10.1051/0004-6361:20020269},
       adsurl = {https://ui.adsabs.harvard.edu/abs/2002A&A...387..244G}
}

@ARTICLE{Schonberner2024,
       author = {{Sch{\"o}nberner}, D. and {Steffen}, M.},
        title = "{Hot bubbles of planetary nebulae with hydrogen-deficient winds. III. Formation and evolution in comparison with hydrogen-rich bubbles}",
      journal = {\aap},
         year = 2024,
        month = apr,
       volume = {684},
          eid = {A105},
        pages = {A105},
          doi = {10.1051/0004-6361/202346170},
archivePrefix = {arXiv},
       eprint = {2401.16309},
 primaryClass = {astro-ph.SR},
       adsurl = {https://ui.adsabs.harvard.edu/abs/2024A&A...684A.105S}
}

@ARTICLE{MB2006,
       author = {{Miller Bertolami}, M.~M. and {Althaus}, L.~G.},
        title = "{Full evolutionary models for PG 1159 stars. Implications for the helium-rich O(He) stars}",
      journal = {\aap},
         year = 2006,
        month = aug,
       volume = {454},
       number = {3},
        pages = {845-854},
          doi = {10.1051/0004-6361:20054723},
archivePrefix = {arXiv},
       eprint = {astro-ph/0603846},
 primaryClass = {astro-ph},
       adsurl = {https://ui.adsabs.harvard.edu/abs/2006A&A...454..845M}
}

@ARTICLE{Wesson2003,
       author = {{Wesson}, R. and {Liu}, X.-W. and {Barlow}, M.~J.},
        title = "{Physical conditions in the planetary nebula Abell 30}",
      journal = {\mnras},
         year = 2003,
        month = mar,
       volume = {340},
       number = {1},
        pages = {253-263},
          doi = {10.1046/j.1365-8711.2003.06289.x},
archivePrefix = {arXiv},
       eprint = {astro-ph/0301119},
 primaryClass = {astro-ph},
       adsurl = {https://ui.adsabs.harvard.edu/abs/2003MNRAS.340..253W}
}

@ARTICLE{Lau2011,
       author = {{Lau}, Herbert H.~B. and {De Marco}, Orsola and {Liu}, X.-W.},
        title = "{V605 Aquilae: a born-again star, a nova or both?}",
      journal = {\mnras},
         year = 2011,
        month = jan,
       volume = {410},
       number = {3},
        pages = {1870-1876},
          doi = {10.1111/j.1365-2966.2010.17568.x},
archivePrefix = {arXiv},
       eprint = {1009.3138},
 primaryClass = {astro-ph.SR},
       adsurl = {https://ui.adsabs.harvard.edu/abs/2011MNRAS.410.1870L}
}

@ARTICLE{Verbena2011,
       author = {{Verbena}, J.~L. and {Schr{\"o}der}, K. -P. and {Wachter}, A.},
        title = "{Outflow dynamics of dust-driven wind models and implications for cool envelopes of PNe}",
      journal = {\mnras},
         year = 2011,
        month = aug,
       volume = {415},
       number = {3},
        pages = {2270-2274},
          doi = {10.1111/j.1365-2966.2011.18859.x},
archivePrefix = {arXiv},
       eprint = {1106.5093},
 primaryClass = {astro-ph.SR},
       adsurl = {https://ui.adsabs.harvard.edu/abs/2011MNRAS.415.2270V}
}

@ARTICLE{Marigo2009,
       author = {{Marigo}, P. and {Aringer}, B.},
        title = "{Low-temperature gas opacity. {\AE}SOPUS: a versatile and quick computational tool}",
      journal = {\aap},
         year = 2009,
        month = dec,
       volume = {508},
       number = {3},
        pages = {1539-1569},
          doi = {10.1051/0004-6361/200912598},
archivePrefix = {arXiv},
       eprint = {0907.3248},
 primaryClass = {astro-ph.SR},
       adsurl = {https://ui.adsabs.harvard.edu/abs/2009A&A...508.1539M}
}

@ARTICLE{Iglesias1993,
       author = {{Iglesias}, Carlos A. and {Rogers}, Forrest J.},
        title = "{Radiative Opacities for Carbon- and Oxygen-rich Mixtures}",
      journal = {\apj},
         year = 1993,
        month = aug,
       volume = {412},
        pages = {752},
          doi = {10.1086/172958},
       adsurl = {https://ui.adsabs.harvard.edu/abs/1993ApJ...412..752I}
}

@ARTICLE{Iglesias1996,
       author = {{Iglesias}, Carlos A. and {Rogers}, Forrest J.},
        title = "{Updated Opal Opacities}",
      journal = {\apj},
         year = 1996,
        month = jun,
       volume = {464},
        pages = {943},
          doi = {10.1086/177381},
       adsurl = {https://ui.adsabs.harvard.edu/abs/1996ApJ...464..943I}
}

@ARTICLE{vanMarle2012,
       author = {{van Marle}, A.~J. and {Keppens}, R.},
        title = "{Multi-dimensional models of circumstellar shells around evolved massive stars}",
      journal = {\aap},
         year = 2012,
        month = nov,
       volume = {547},
          eid = {A3},
        pages = {A3},
          doi = {10.1051/0004-6361/201218957},
archivePrefix = {arXiv},
       eprint = {1209.4496},
 primaryClass = {astro-ph.SR},
       adsurl = {https://ui.adsabs.harvard.edu/abs/2012A&A...547A...3V}
}

@ARTICLE{Sander2024,
       author = {{Sander}, A.~A.~C. and {Bouret}, J. -C. and {Bernini-Peron}, M. and {Puls}, J. and {Backs}, F. and {Berlanas}, S.~R. and {Bestenlehner}, J.~M. and {Brands}, S.~A. and {Herrero}, A. and {Martins}, F. and {Maryeva}, O. and {Pauli}, D. and {Ramachandran}, V. and {Crowther}, P.~A. and {G{\'o}mez-Gonz{\'a}lez}, V.~M.~A. and {Gormaz-Matamala}, A.~C. and {Hamann}, W. -R. and {Hillier}, D.~J. and {Kuiper}, R. and {Larkin}, C.~J.~K. and {Lefever}, R.~R. and {Mehner}, A. and {Najarro}, F. and {Oskinova}, L.~M. and {Sch{\"o}sser}, E.~C. and {Shenar}, T. and {Todt}, H. and {ud-Doula}, A. and {Vink}, J.~S.},
        title = "{X-Shooting ULLYSES: Massive stars at low metallicity: IV. Spectral analysis methods and exemplary results for O stars}",
      journal = {\aap},
         year = 2024,
        month = sep,
       volume = {689},
          eid = {A30},
        pages = {A30},
          doi = {10.1051/0004-6361/202449829},
archivePrefix = {arXiv},
       eprint = {2407.03137},
 primaryClass = {astro-ph.SR},
       adsurl = {https://ui.adsabs.harvard.edu/abs/2024A&A...689A..30S}
}

@ARTICLE{Jeffery2006,
       author = {{Jeffery}, C.~S. and {Sch{\"o}nberner}, D.},
        title = "{Stellar archaeology: the evolving spectrum of FG Sagittae}",
      journal = {\aap},
         year = 2006,
        month = dec,
       volume = {459},
       number = {3},
        pages = {885-899},
          doi = {10.1051/0004-6361:20047075},
archivePrefix = {arXiv},
       eprint = {astro-ph/0608542},
 primaryClass = {astro-ph},
       adsurl = {https://ui.adsabs.harvard.edu/abs/2006A&A...459..885J}
}

@ARTICLE{Reindl2017,
       author = {{Reindl}, Nicole and {Rauch}, T. and {Miller Bertolami}, M.~M. and {Todt}, H. and {Werner}, K.},
        title = "{Breaking news from the HST: the central star of the Stingray Nebula is now returning towards the AGB}",
      journal = {\mnras},
         year = 2017,
        month = jan,
       volume = {464},
       number = {1},
        pages = {L51-L55},
          doi = {10.1093/mnrasl/slw175},
archivePrefix = {arXiv},
       eprint = {1609.07113},
 primaryClass = {astro-ph.SR},
       adsurl = {https://ui.adsabs.harvard.edu/abs/2017MNRAS.464L..51R}
}

@ARTICLE{Lawlor2003,
       author = {{Lawlor}, T.~M. and {MacDonald}, J.},
        title = "{Sakurai's Object, V605 Aquilae, and FG Sagittae: An Evolutionary Sequence Revealed}",
      journal = {\apj},
         year = 2003,
        month = feb,
       volume = {583},
       number = {2},
        pages = {913-922},
          doi = {10.1086/345411},
       adsurl = {https://ui.adsabs.harvard.edu/abs/2003ApJ...583..913L}
}

@ARTICLE{Herwig1999,
       author = {{Herwig}, F. and {Bl{\"o}cker}, T. and {Langer}, N. and {Driebe}, T.},
        title = "{On the formation of hydrogen-deficient post-AGB stars}",
      journal = {\aap},
         year = 1999,
        month = sep,
       volume = {349},
        pages = {L5-L8},
          doi = {10.48550/arXiv.astro-ph/9908108},
archivePrefix = {arXiv},
       eprint = {astro-ph/9908108},
 primaryClass = {astro-ph},
       adsurl = {https://ui.adsabs.harvard.edu/abs/1999A&A...349L...5H}
}

@ARTICLE{MB2005,
       author = {{Miller Bertolami}, M.~M. and {Alhaus}, L.~G. and {C{\'o}rsico}, A.~H.},
        title = "{The formation of DA white dwarfs with thin Hydrogen envelopes through a late thermal pulse}",
      journal = {Boletin de la Asociacion Argentina de Astronomia La Plata Argentina},
         year = 2005,
        month = jan,
       volume = {48},
        pages = {185-191},
       adsurl = {https://ui.adsabs.harvard.edu/abs/2005BAAA...48..185M}
}

@ARTICLE{Rubio2022,
       author = {{Rubio}, G. and {Toal{\'a}}, J.~A. and {Todt}, H. and {Sabin}, L. and {Santamar{\'\i}a}, E. and {Ramos-Larios}, G. and {Guerrero}, M.~A.},
        title = "{Planetary nebulae with Wolf-Rayet-type central stars - IV. NGC 1501 and its mixing layer}",
      journal = {\mnras},
         year = 2022,
        month = dec,
       volume = {517},
       number = {4},
        pages = {5166-5179},
          doi = {10.1093/mnras/stac3011},
archivePrefix = {arXiv},
       eprint = {2210.09116},
 primaryClass = {astro-ph.SR},
       adsurl = {https://ui.adsabs.harvard.edu/abs/2022MNRAS.517.5166R}
}

@ARTICLE{GG2020,
       author = {{G{\'o}mez-Gonz{\'a}lez}, V.~M.~A. and {Toal{\'a}}, J.~A. and {Guerrero}, M.~A. and {Todt}, H. and {Sabin}, L. and {Ramos-Larios}, G. and {Mayya}, Y.~D.},
        title = "{Planetary nebulae with Wolf-Rayet-type central stars - I. The case of the high-excitation NGC 2371}",
      journal = {\mnras},
         year = 2020,
        month = jul,
       volume = {496},
       number = {1},
        pages = {959-973},
          doi = {10.1093/mnras/staa1542},
archivePrefix = {arXiv},
       eprint = {2005.14294},
 primaryClass = {astro-ph.SR},
       adsurl = {https://ui.adsabs.harvard.edu/abs/2020MNRAS.496..959G}
}

@ARTICLE{GG2022,
       author = {{G{\'o}mez-Gonz{\'a}lez}, V.~M.~A. and {Rubio}, G. and {Toal{\'a}}, J.~A. and {Guerrero}, M.~A. and {Sabin}, L. and {Todt}, H. and {G{\'o}mez-Llanos}, V. and {Ramos-Larios}, G. and {Mayya}, Y.~D.},
        title = "{Planetary nebulae with Wolf-Rayet-type central stars - III. A detailed view of NGC 6905 and its central star}",
      journal = {\mnras},
         year = 2022,
        month = jan,
       volume = {509},
       number = {1},
        pages = {974-989},
          doi = {10.1093/mnras/stab3042},
archivePrefix = {arXiv},
       eprint = {2110.09551},
 primaryClass = {astro-ph.SR},
       adsurl = {https://ui.adsabs.harvard.edu/abs/2022MNRAS.509..974G}
}

@ARTICLE{Meaburn1996,
       author = {{Meaburn}, J. and {Lopez}, J.~A.},
        title = "{The Dramatic Kinematics of the Hydrogen-Deficient Planetary Nebula Abell 30}",
      journal = {\apjl},
         year = 1996,
        month = nov,
       volume = {472},
        pages = {L45},
          doi = {10.1086/310357},
       adsurl = {https://ui.adsabs.harvard.edu/abs/1996ApJ...472L..45M}
}

@ARTICLE{Meaburn1998,
       author = {{Meaburn}, J. and {Lopez}, J.~A. and {Bryce}, M. and {Redman}, M.~P.},
        title = "{Localised, high-speed flows within the hydrogen-deficient planetary nebula Abell 78}",
      journal = {\aap},
         year = 1998,
        month = jun,
       volume = {334},
        pages = {670-677},
       adsurl = {https://ui.adsabs.harvard.edu/abs/1998A&A...334..670M}
}

@ARTICLE{Kudritzki1989,
       author = {{Kudritzki}, R.~P. and {Pauldrach}, A. and {Puls}, J. and {Abbott}, D.~C.},
        title = "{Radiation-driven winds of hot stars. VI. Analytical solutions for wind models including the finite cone angle effect.}",
      journal = {\aap},
         year = 1989,
        month = jul,
       volume = {219},
        pages = {205-218},
       adsurl = {https://ui.adsabs.harvard.edu/abs/1989A&A...219..205K}
}

@ARTICLE{Soker1997,
       author = {{Soker}, Noam},
        title = "{Properties That Cannot Be Explained by the Progenitors of Planetary Nebulae}",
      journal = {\apjs},
         year = 1997,
        month = oct,
       volume = {112},
       number = {2},
        pages = {487-505},
          doi = {10.1086/313040},
       adsurl = {https://ui.adsabs.harvard.edu/abs/1997ApJS..112..487S}
}

@ARTICLE{RodriguezGonzalez2022,
       author = {{Rodr{\'\i}guez-Gonz{\'a}lez}, J.~B. and {Santamar{\'\i}a}, E. and {Toal{\'a}}, J.~A. and {Guerrero}, M.~A. and {Montoro-Molina}, B. and {Rubio}, G. and {Tafoya}, D. and {Chu}, Y. -H. and {Ramos-Larios}, G. and {Sabin}, L.},
        title = "{Common envelope evolution in born-again planetary nebulae - Shaping the H-deficient ejecta of A 30}",
      journal = {\mnras},
         year = 2022,
        month = aug,
       volume = {514},
       number = {4},
        pages = {4794-4802},
          doi = {10.1093/mnras/stac1697},
archivePrefix = {arXiv},
       eprint = {2204.08501},
 primaryClass = {astro-ph.SR},
       adsurl = {https://ui.adsabs.harvard.edu/abs/2022MNRAS.514.4794R}
}

@ARTICLE{Ary2023,
       author = {{Rodr{\'\i}guez-Gonz{\'a}lez}, Ary and {Pe{\~n}a}, Miriam and {Hern{\'a}ndez-Mart{\'\i}nez}, Liliana and {Ruiz-Escobedo}, Francisco and {Raga}, Alejandro and {Stasi{\'n}ska}, Grazyna and {Castorena}, Jorge Ivan},
        title = "{Numerical Models of Planetary Nebulae with Different Episodes of Mass Ejection: The Particular Case of HuBi 1}",
      journal = {\apj},
         year = 2023,
        month = oct,
       volume = {955},
       number = {2},
          eid = {151},
        pages = {151},
          doi = {10.3847/1538-4357/acf0bc},
archivePrefix = {arXiv},
       eprint = {2308.13190},
 primaryClass = {astro-ph.GA},
       adsurl = {https://ui.adsabs.harvard.edu/abs/2023ApJ...955..151R}
}

@ARTICLE{RechyGarcia2020,
       author = {{Rechy-Garc{\'\i}a}, J.~S. and {Guerrero}, M.~A. and {Santamar{\'\i}a}, E. and {G{\'o}mez-Gonz{\'a}lez}, V.~M.~A. and {Ramos-Larios}, G. and {Toal{\'a}}, J.~A. and {Cazzoli}, S. and {Sabin}, L. and {Miranda}, L.~F. and {Fang}, X. and {Liu}, J.},
        title = "{Discovery of a Fast-expanding Shell in the Inside-out Born-again Planetary Nebula HuBi 1 through High-dispersion Integral Field Spectroscopy}",
      journal = {\apjl},
         year = 2020,
        month = nov,
       volume = {903},
       number = {1},
          eid = {L4},
        pages = {L4},
          doi = {10.3847/2041-8213/abbe22},
archivePrefix = {arXiv},
       eprint = {2009.13575},
 primaryClass = {astro-ph.SR},
       adsurl = {https://ui.adsabs.harvard.edu/abs/2020ApJ...903L...4R}
}

@ARTICLE{Guerrero2018,
       author = {{Guerrero}, Mart{\'\i}n A. and {Fang}, Xuan and {Miller Bertolami}, Marcelo M. and {Ramos-Larios}, Gerardo and {Todt}, Helge and {Alarie}, Alexandre and {Sabin}, Laurence and {Miranda}, Luis F. and {Morisset}, Christophe and {Kehrig}, Carolina and {Zavala}, Sa{\'u}l A.},
        title = "{The inside-out planetary nebula around a born-again star}",
      journal = {Nature Astronomy},
         year = 2018,
        month = aug,
       volume = {2},
        pages = {784-789},
          doi = {10.1038/s41550-018-0551-8},
archivePrefix = {arXiv},
       eprint = {1808.03462},
 primaryClass = {astro-ph.SR},
       adsurl = {https://ui.adsabs.harvard.edu/abs/2018NatAs...2..784G}
}

@ARTICLE{Clayton2013,
       author = {{Clayton}, Geoffrey C. and {Bond}, Howard E. and {Long}, Lindsey A. and {Meyer}, Paul I. and {Sugerman}, Ben E.~K. and {Montiel}, Edward and {Sparks}, William B. and {Meakes}, M.~G. and {Chesneau}, O. and {De Marco}, O.},
        title = "{Evolution of the 1919 Ejecta of V605 Aquilae}",
      journal = {\apj},
         year = 2013,
        month = jul,
       volume = {771},
       number = {2},
          eid = {130},
        pages = {130},
          doi = {10.1088/0004-637X/771/2/130},
archivePrefix = {arXiv},
       eprint = {1305.6563},
 primaryClass = {astro-ph.SR},
       adsurl = {https://ui.adsabs.harvard.edu/abs/2013ApJ...771..130C}
}

@ARTICLE{MontoroMolina2022,
       author = {{Montoro-Molina}, B. and {Guerrero}, M.~A. and {P{\'e}rez-D{\'\i}az}, B. and {Toal{\'a}}, J.~A. and {Cazzoli}, S. and {Miller Bertolami}, M.~M. and {Morisset}, C.},
        title = "{Chemistry and physical properties of the born-again planetary nebula HuBi 1}",
      journal = {\mnras},
         year = 2022,
        month = may,
       volume = {512},
       number = {3},
        pages = {4003-4020},
          doi = {10.1093/mnras/stac336},
archivePrefix = {arXiv},
       eprint = {2202.00353},
 primaryClass = {astro-ph.SR},
       adsurl = {https://ui.adsabs.harvard.edu/abs/2022MNRAS.512.4003M}
}

@ARTICLE{MontoroMolina2024,
       author = {{Montoro-Molina}, B. and {Tafoya}, D. and {Guerrero}, M.~A. and {Toal{\'a}}, J.~A. and {Santamar{\'\i}a}, E.},
        title = "{Optical tomography of the born-again ejecta of A 58}",
      journal = {\aap},
         year = 2024,
        month = apr,
       volume = {684},
          eid = {A107},
        pages = {A107},
          doi = {10.1051/0004-6361/202348528},
archivePrefix = {arXiv},
       eprint = {2401.09844},
 primaryClass = {astro-ph.SR},
       adsurl = {https://ui.adsabs.harvard.edu/abs/2024A&A...684A.107M}
}

@ARTICLE{Fang2014,
       author = {{Fang}, X. and {Guerrero}, M.~A. and {Marquez-Lugo}, R.~A. and {Toal{\'a}}, J.~A. and {Arthur}, S.~J. and {Chu}, Y. -H. and {Blair}, W.~P. and {Gruendl}, R.~A. and {Hamann}, W. -R. and {Oskinova}, L.~M. and {Todt}, H.},
        title = "{Expansion of Hydrogen-poor Knots in the Born-again Planetary Nebulae A30 and A78}",
      journal = {\apj},
         year = 2014,
        month = dec,
       volume = {797},
       number = {2},
          eid = {100},
        pages = {100},
          doi = {10.1088/0004-637X/797/2/100},
archivePrefix = {arXiv},
       eprint = {1410.3872},
 primaryClass = {astro-ph.SR},
       adsurl = {https://ui.adsabs.harvard.edu/abs/2014ApJ...797..100F}
}

@ARTICLE{Paczynski1967,
       author = {{Paczy{\'n}ski}, B.},
        title = "{Evolution of Close Binaries. V. The Evolution of Massive Binaries and the Formation of the Wolf-Rayet Stars}",
      journal = {\actaa},
         year = 1967,
        month = jan,
       volume = {17},
        pages = {355},
       adsurl = {https://ui.adsabs.harvard.edu/abs/1967AcA....17..355P}
}

@ARTICLE{Paczynski1971,
       author = {{Paczy{\'n}ski}, B.},
        title = "{Evolution of Single Stars. VI. Model Nuclei of Planetary Nebulae}",
      journal = {\actaa},
         year = 1971,
        month = jan,
       volume = {21},
        pages = {417},
       adsurl = {https://ui.adsabs.harvard.edu/abs/1971AcA....21..417P}
}

@ARTICLE{Schwab2019,
       author = {{Schwab}, Josiah},
        title = "{Evolutionary Models for R Coronae Borealis Stars}",
      journal = {\apj},
         year = 2019,
        month = nov,
       volume = {885},
       number = {1},
          eid = {27},
        pages = {27},
          doi = {10.3847/1538-4357/ab425d},
archivePrefix = {arXiv},
       eprint = {1909.02569},
 primaryClass = {astro-ph.SR},
       adsurl = {https://ui.adsabs.harvard.edu/abs/2019ApJ...885...27S}
}

@ARTICLE{Freeman2014,
       author = {{Freeman}, M. and {Montez}, Jr., R. and {Kastner}, J.~H. and {Balick}, B. and {Frew}, D.~J. and {Jones}, D. and {Miszalski}, B. and {Sahai}, R. and {Blackman}, E. and {Chu}, Y. -H. and {De Marco}, O. and {Frank}, A. and {Guerrero}, M.~A. and {Lopez}, J.~A. and {Zijlstra}, A. and {Bujarrabal}, V. and {Corradi}, R.~L.~M. and {Nordhaus}, J. and {Parker}, Q.~A. and {Sandin}, C. and {Sch{\"o}nberner}, D. and {Soker}, N. and {Sokoloski}, J.~L. and {Steffen}, M. and {Toal{\'a}}, J.~A. and {Ueta}, T. and {Villaver}, E.},
        title = "{The Chandra Planetary Nebula Survey (ChanPlaNS). II. X-Ray Emission from Compact Planetary Nebulae}",
      journal = {\apj},
         year = 2014,
        month = oct,
       volume = {794},
       number = {2},
          eid = {99},
        pages = {99},
          doi = {10.1088/0004-637X/794/2/99},
archivePrefix = {arXiv},
       eprint = {1407.4141},
 primaryClass = {astro-ph.SR},
       adsurl = {https://ui.adsabs.harvard.edu/abs/2014ApJ...794...99F}
}

@ARTICLE{Steffen2008,
       author = {{Steffen}, M. and {Sch{\"o}nberner}, D. and {Warmuth}, A.},
        title = "{The evolution of planetary nebulae. V. The diffuse X-ray emission}",
      journal = {\aap},
         year = 2008,
        month = oct,
       volume = {489},
       number = {1},
        pages = {173-194},
          doi = {10.1051/0004-6361:200809677},
archivePrefix = {arXiv},
       eprint = {0807.3290},
 primaryClass = {astro-ph},
       adsurl = {https://ui.adsabs.harvard.edu/abs/2008A&A...489..173S}
}

@ARTICLE{Schonberner1979,
       author = {{Schönberner}, D.},
        title = "{Asymptotic giant branch evolution with steady mass loss.}",
      journal = {\aap},
         year = 1979,
        month = oct,
       volume = {79},
        pages = {108-114},
       adsurl = {https://ui.adsabs.harvard.edu/abs/1979A&A....79..108S}
}

@ARTICLE{Blocker2001,
       author = {{Bl{\"o}cker}, Thomas},
        title = "{Evolution on the AGB and beyond: on the formation of H-deficient post-AGB stars}",
      journal = {\apss},
         year = 2001,
        month = jan,
       volume = {275},
        pages = {1-14},
          doi = {10.1023/A:1002777931450},
archivePrefix = {arXiv},
       eprint = {astro-ph/0102135},
 primaryClass = {astro-ph},
       adsurl = {https://ui.adsabs.harvard.edu/abs/2001Ap&SS.275....1B}
}

@ARTICLE{Aller1983,
       author = {{Aller}, L.~H. and {Czyzak}, S.~J.},
        title = "{Chemical compositions of planetary nebulae.}",
      journal = {\apjs},
         year = 1983,
        month = feb,
       volume = {51},
        pages = {211-248},
          doi = {10.1086/190846},
       adsurl = {https://ui.adsabs.harvard.edu/abs/1983ApJS...51..211A}
}

@ARTICLE{Toala2018,
       author = {{Toal{\'a}}, J.~A. and {Arthur}, S.~J.},
        title = "{On the X-ray temperature of hot gas in diffuse nebulae}",
      journal = {\mnras},
         year = 2018,
        month = jul,
       volume = {478},
       number = {1},
        pages = {1218-1230},
          doi = {10.1093/mnras/sty1127},
archivePrefix = {arXiv},
       eprint = {1804.10244},
 primaryClass = {astro-ph.GA},
       adsurl = {https://ui.adsabs.harvard.edu/abs/2018MNRAS.478.1218T}
}

@ARTICLE{Tafoya2022,
       author = {{Tafoya}, Daniel and {Toal{\'a}}, Jes{\'u}s A. and {Unnikrishnan}, Ramlal and {Vlemmings}, Wouter H.~T. and {Guerrero}, Mart{\'\i}n A. and {Kimeswenger}, Stefan and {van Hoof}, Peter A.~M. and {Zapata}, Luis A. and {Trevi{\~n}o-Morales}, Sandra P. and {Rodr{\'\i}guez-Gonz{\'a}lez}, Janis B.},
        title = "{First Images of the Molecular Gas around a Born-again Star Revealed by ALMA}",
      journal = {\apjl},
         year = 2022,
        month = jan,
       volume = {925},
       number = {1},
          eid = {L4},
        pages = {L4},
          doi = {10.3847/2041-8213/ac4a5b},
archivePrefix = {arXiv},
       eprint = {2201.04110},
 primaryClass = {astro-ph.GA},
       adsurl = {https://ui.adsabs.harvard.edu/abs/2022ApJ...925L...4T}
}

@ARTICLE{Tafoya2023,
       author = {{Tafoya}, Daniel and {van Hoof}, Peter A.~M. and {Toal{\'a}}, Jes{\'u}s A. and {Van de Steene}, Griet and {Randall}, Suzanna and {Unnikrishnan}, Ramlal and {Kimeswenger}, Stefan and {Hajduk}, Marcin and {Barr{\'\i}a}, Daniela and {Zijlstra}, Albert},
        title = "{The heart of Sakurai's object revealed by ALMA}",
      journal = {\aap},
         year = 2023,
        month = sep,
       volume = {677},
          eid = {L8},
        pages = {L8},
          doi = {10.1051/0004-6361/202347293},
archivePrefix = {arXiv},
       eprint = {2308.08962},
 primaryClass = {astro-ph.SR},
       adsurl = {https://ui.adsabs.harvard.edu/abs/2023A&A...677L...8T}
}

@ARTICLE{Nakano1996,
       author = {{Nakano}, S. and {Sakurai}, Y. and {Hazen}, M. and {McNaught}, R.~H. and {Benetti}, S. and {Duerbeck}, H.~W. and {Cappellaro}, E. and {Leibundgut}, B.},
        title = "{Novalike Variable in Sagittarius}",
      journal = {\iaucirc},
         year = 1996,
        month = feb,
       volume = {6322},
        pages = {1},
       adsurl = {https://ui.adsabs.harvard.edu/abs/1996IAUC.6322....1N}
}

@ARTICLE{Clayton1997,
       author = {{Clayton}, Geoffrey C. and {De Marco}, Orsola},
        title = "{The Evolution of the Final Helium Shell Flash Star V605 Aquilae From 1917 to 1997}",
      journal = {\aj},
         year = 1997,
        month = dec,
       volume = {114},
        pages = {2679},
          doi = {10.1086/118678},
       adsurl = {https://ui.adsabs.harvard.edu/abs/1997AJ....114.2679C}
}

@INPROCEEDINGS{Todt2015,
       author = {{Todt}, H. and {Hamann}, W. -R.},
        title = "{Wolf-Rayet central stars of planetary nebulae}",
    booktitle = {Wolf-Rayet Stars},
         year = 2015,
       editor = {{Hamann}, Wolf-Rainer and {Sander}, Andreas and {Todt}, Helge},
        month = jan,
        pages = {253-258},
       adsurl = {https://ui.adsabs.harvard.edu/abs/2015wrs..conf..253T}
}

@ARTICLE{Kuiper2020,
       author = {{Kuiper}, Rolf and {Yorke}, Harold W. and {Mignone}, Andrea},
        title = "{Makemake + Sedna: A Continuum Radiation Transport and Photoionization Framework for Astrophysical Newtonian Fluid Dynamics}",
      journal = {\apjs},
         year = 2020,
        month = sep,
       volume = {250},
       number = {1},
          eid = {13},
        pages = {13},
          doi = {10.3847/1538-4365/ab9a36},
archivePrefix = {arXiv},
       eprint = {2009.12374},
 primaryClass = {astro-ph.IM},
       adsurl = {https://ui.adsabs.harvard.edu/abs/2020ApJS..250...13K}
}

@ARTICLE{Mignone2007,
       author = {{Mignone}, A. and {Bodo}, G. and {Massaglia}, S. and {Matsakos}, T. and {Tesileanu}, O. and {Zanni}, C. and {Ferrari}, A.},
        title = "{PLUTO: A Numerical Code for Computational Astrophysics}",
      journal = {\apjs},
         year = 2007,
        month = may,
       volume = {170},
       number = {1},
        pages = {228-242},
          doi = {10.1086/513316},
archivePrefix = {arXiv},
       eprint = {astro-ph/0701854},
 primaryClass = {astro-ph},
       adsurl = {https://ui.adsabs.harvard.edu/abs/2007ApJS..170..228M}
}

@BOOK{Dyson1997,
       author = {{Dyson}, J.~E. and {Williams}, D.~A.},
        title = "{The physics of the interstellar medium}",
         year = 1997,
          doi = {10.1201/9780585368115},
       adsurl = {https://ui.adsabs.harvard.edu/abs/1997pism.book.....D}
}

@ARTICLE{Iben1983,
       author = {{Iben}, Jr., I. and {Kaler}, J.~B. and {Truran}, J.~W. and {Renzini}, A.},
        title = "{On the evolution of those nuclei of planetary nebulae that experiencea final helium shell flash.}",
      journal = {\apj},
         year = 1983,
        month = jan,
       volume = {264},
        pages = {605-612},
          doi = {10.1086/160631},
       adsurl = {https://ui.adsabs.harvard.edu/abs/1983ApJ...264..605I}
}

@ARTICLE{MB2024,
       author = {{Miller Bertolami}, Marcelo M.},
        title = "{Primer on Formation and Evolution of Hydrogen-Deficient Central Stars of Planetary Nebul{\ae} and Related Objects}",
      journal = {Galaxies},
         year = 2024,
        month = nov,
       volume = {12},
       number = {6},
          eid = {83},
        pages = {83},
          doi = {10.3390/galaxies12060083},
archivePrefix = {arXiv},
       eprint = {2411.18035},
 primaryClass = {astro-ph.SR},
       adsurl = {https://ui.adsabs.harvard.edu/abs/2024Galax..12...83M}
}

@ARTICLE{Ramstedt2020,
       author = {{Ramstedt}, S. and {Vlemmings}, W.~H.~T. and {Doan}, L. and {Danilovich}, T. and {Lindqvist}, M. and {Saberi}, M. and {Olofsson}, H. and {De Beck}, E. and {Groenewegen}, M.~A.~T. and {H{\"o}fner}, S. and {Kastner}, J.~H. and {Kerschbaum}, F. and {Khouri}, T. and {Maercker}, M. and {Montez}, R. and {Quintana-Lacaci}, G. and {Sahai}, R. and {Tafoya}, D. and {Zijlstra}, A.},
        title = "{DEATHSTAR: Nearby AGB stars with the Atacama Compact Array. I. CO envelope sizes and asymmetries: A new hope for accurate mass-loss-rate estimates}",
      journal = {\aap},
         year = 2020,
        month = aug,
       volume = {640},
          eid = {A133},
        pages = {A133},
          doi = {10.1051/0004-6361/201936874},
archivePrefix = {arXiv},
       eprint = {2008.07885},
 primaryClass = {astro-ph.SR},
       adsurl = {https://ui.adsabs.harvard.edu/abs/2020A&A...640A.133R}
}

@ARTICLE{Stute2006,
       author = {{Stute}, Matthias and {Sahai}, Raghvendra},
        title = "{X-Ray Emission from Planetary Nebulae. I. Spherically Symmetric Numerical Simulations}",
      journal = {\apj},
         year = 2006,
        month = nov,
       volume = {651},
       number = {2},
        pages = {882-897},
          doi = {10.1086/507986},
archivePrefix = {arXiv},
       eprint = {astro-ph/0609442},
 primaryClass = {astro-ph},
       adsurl = {https://ui.adsabs.harvard.edu/abs/2006ApJ...651..882S}
}

@ARTICLE{Toala2016,
       author = {{Toal{\'a}}, J.~A. and {Arthur}, S.~J.},
        title = "{Formation and X-ray emission from hot bubbles in planetary nebulae - II. Hot bubble X-ray emission}",
      journal = {\mnras},
         year = 2016,
        month = dec,
       volume = {463},
       number = {4},
        pages = {4438-4458},
          doi = {10.1093/mnras/stw2307},
archivePrefix = {arXiv},
       eprint = {1609.03197},
 primaryClass = {astro-ph.SR},
       adsurl = {https://ui.adsabs.harvard.edu/abs/2016MNRAS.463.4438T}
}

@ARTICLE{Toala2014,
       author = {{Toal{\'a}}, J.~A. and {Arthur}, S.~J.},
        title = "{Formation and X-ray emission from hot bubbles in planetary nebulae - I. Hot bubble formation}",
      journal = {\mnras},
         year = 2014,
        month = oct,
       volume = {443},
       number = {4},
        pages = {3486-3505},
          doi = {10.1093/mnras/stu1360},
archivePrefix = {arXiv},
       eprint = {1407.1421},
 primaryClass = {astro-ph.SR},
       adsurl = {https://ui.adsabs.harvard.edu/abs/2014MNRAS.443.3486T}
}

@ARTICLE{Arthur2006,
       author = {{Arthur}, S. Jane and {Hoare}, M.~G.},
        title = "{Hydrodynamics of Cometary Compact H II Regions}",
      journal = {\apjs},
         year = 2006,
        month = jul,
       volume = {165},
       number = {1},
        pages = {283-306},
          doi = {10.1086/503899},
archivePrefix = {arXiv},
       eprint = {astro-ph/0511035},
 primaryClass = {astro-ph},
       adsurl = {https://ui.adsabs.harvard.edu/abs/2006ApJS..165..283A}
}

@ARTICLE{Toala2011,
       author = {{Toal{\'a}}, J.~A. and {Arthur}, S.~J.},
        title = "{Radiation-hydrodynamic Models of the Evolving Circumstellar Medium around Massive Stars}",
      journal = {\apj},
         year = 2011,
        month = aug,
       volume = {737},
       number = {2},
          eid = {100},
        pages = {100},
          doi = {10.1088/0004-637X/737/2/100},
archivePrefix = {arXiv},
       eprint = {1106.4493},
 primaryClass = {astro-ph.SR},
       adsurl = {https://ui.adsabs.harvard.edu/abs/2011ApJ...737..100T}
}

@ARTICLE{Villaver2002a,
       author = {{Villaver}, Eva and {Garc{\'\i}a-Segura}, Guillermo and {Manchado}, Arturo},
        title = "{The Dynamical Evolution of the Circumstellar Gas around Low- and Intermediate-Mass Stars. I. The Asymptotic Giant Branch}",
      journal = {\apj},
         year = 2002,
        month = jun,
       volume = {571},
       number = {2},
        pages = {880-900},
          doi = {10.1086/340022},
archivePrefix = {arXiv},
       eprint = {astro-ph/0202050},
 primaryClass = {astro-ph},
       adsurl = {https://ui.adsabs.harvard.edu/abs/2002ApJ...571..880V}
}

@ARTICLE{vanderHucht1981,
       author = {{van der Hucht}, K.~A. and {Conti}, P.~S. and {Lundstrom}, I. and {Stenholm}, B.},
        title = "{The Sixth Catalogue of Galactic Wolf-Rayet Stars - Their Past and Present}",
      journal = {\ssr},
         year = 1981,
        month = sep,
       volume = {28},
       number = {3},
        pages = {227-306},
          doi = {10.1007/BF00173260},
       adsurl = {https://ui.adsabs.harvard.edu/abs/1981SSRv...28..227V}
}

@ARTICLE{Crowther1998,
       author = {{Crowther}, P.~A. and {De Marco}, Orsola and {Barlow}, M.~J.},
        title = "{Quantitative classification of WC and WO stars}",
      journal = {\mnras},
         year = 1998,
        month = may,
       volume = {296},
       number = {2},
        pages = {367-378},
          doi = {10.1046/j.1365-8711.1998.01360.x},
       adsurl = {https://ui.adsabs.harvard.edu/abs/1998MNRAS.296..367C}
}

@ARTICLE{Acker2003,
       author = {{Acker}, A. and {Neiner}, C.},
        title = "{Quantitative classification of WR nuclei of planetary nebulae}",
      journal = {\aap},
         year = 2003,
        month = may,
       volume = {403},
        pages = {659-673},
          doi = {10.1051/0004-6361:20030391},
       adsurl = {https://ui.adsabs.harvard.edu/abs/2003A&A...403..659A}
}

@ARTICLE{Weidmann2020,
       author = {{Weidmann}, W.~A. and {Mari}, M.~B. and {Schmidt}, E.~O. and {Gaspar}, G. and {Miller Bertolami}, M.~M. and {Oio}, G.~A. and {Guti{\'e}rrez-Soto}, L.~A. and {Volpe}, M.~G. and {Gamen}, R. and {Mast}, D.},
        title = "{Catalogue of the central stars of planetary nebulae. Expanded edition}",
      journal = {\aap},
         year = 2020,
        month = aug,
       volume = {640},
          eid = {A10},
        pages = {A10},
          doi = {10.1051/0004-6361/202037998},
archivePrefix = {arXiv},
       eprint = {2005.10368},
 primaryClass = {astro-ph.GA},
       adsurl = {https://ui.adsabs.harvard.edu/abs/2020A&A...640A..10W}
}

@ARTICLE{Villaver2002b,
       author = {{Villaver}, Eva and {Manchado}, Arturo and {Garc{\'\i}a-Segura}, Guillermo},
        title = "{The Dynamical Evolution of the Circumstellar Gas around Low- and Intermediate-Mass Stars. II. The Planetary Nebula Formation}",
      journal = {\apj},
         year = 2002,
        month = dec,
       volume = {581},
       number = {2},
        pages = {1204-1224},
          doi = {10.1086/344250},
archivePrefix = {arXiv},
       eprint = {astro-ph/0208323},
 primaryClass = {astro-ph},
       adsurl = {https://ui.adsabs.harvard.edu/abs/2002ApJ...581.1204V}
}

@ARTICLE{Wallstrom2025,
       author = {{Wallstr{\"o}m}, S.~H.~J. and {Scicluna}, P. and {Srinivasan}, S. and {Wouterloot}, J.~G.~A. and {McDonald}, I. and {Decock}, L. and {Wijshoff}, M. and {Chen}, R. and {Torres}, D. and {Umans}, L. and {Willebrords}, B. and {Kemper}, F. and {Rau}, G. and {Feng}, S. and {Jeste}, M. and {Kaminski}, T. and {Li}, D. and {Liu}, F.~C. and {Trejo-Cruz}, A. and {Chawner}, H. and {Goldman}, S. and {MacIsaac}, H. and {Tang}, J. and {Zeegers}, S.~T. and {Danilovich}, T. and {Matsuura}, M. and {Menten}, K.~M. and {van Loon}, J. Th and {Cami}, J. and {Clark}, C.~J.~R. and {Dharmawardena}, T.~E. and {Greaves}, J. and {He}, Jinhua and {Imai}, H. and {Jones}, O.~C. and {Kim}, H. and {Marshall}, J.~P. and {Shinnaga}, H. and {Wesson}, R.},
        title = "{The Nearby Evolved Stars Survey: III. First data release of JCMT CO-line observations}",
      journal = {\aap},
         year = 2025,
        month = dec,
       volume = {704},
          eid = {A276},
        pages = {A276},
          doi = {10.1051/0004-6361/202556298},
archivePrefix = {arXiv},
       eprint = {2510.14809},
 primaryClass = {astro-ph.GA},
       adsurl = {https://ui.adsabs.harvard.edu/abs/2025A&A...704A.276W}
}

@ARTICLE{Marten1991,
       author = {{Marten}, H. and {Sch\"{o}nberner}, D.},
        title = "{On the dynamical evolution of planetary nebulae.}",
      journal = {\aap},
         year = 1991,
        month = aug,
       volume = {248},
        pages = {590-598},
       adsurl = {https://ui.adsabs.harvard.edu/abs/1991A&A...248..590M}
}

@ARTICLE{Zou2020,
       author = {{Zou}, Yangyuxin and {Frank}, Adam and {Chen}, Zhuo and {Reichardt}, Thomas and {De Marco}, Orsola and {Blackman}, Eric G. and {Nordhaus}, Jason and {Balick}, Bruce and {Carroll-Nellenback}, Jonathan and {Chamandy}, Luke and {Liu}, Baowei},
        title = "{Bipolar planetary nebulae from outflow collimation by common envelope evolution}",
      journal = {\mnras},
         year = 2020,
        month = sep,
       volume = {497},
       number = {3},
        pages = {2855-2869},
          doi = {10.1093/mnras/staa2145},
archivePrefix = {arXiv},
       eprint = {1912.01647},
 primaryClass = {astro-ph.SR},
       adsurl = {https://ui.adsabs.harvard.edu/abs/2020MNRAS.497.2855Z}
}

@ARTICLE{Perinotto2004,
       author = {{Perinotto}, M. and {Sch{\"o}nberner}, D. and {Steffen}, M. and {Calonaci}, C.},
        title = "{The evolution of planetary nebulae. I. A radiation-hydrodynamics parameter study}",
      journal = {\aap},
         year = 2004,
        month = feb,
       volume = {414},
        pages = {993-1015},
          doi = {10.1051/0004-6361:20031653},
       adsurl = {https://ui.adsabs.harvard.edu/abs/2004A&A...414..993P}
}

@ARTICLE{Guerrero2013,
       author = {{Guerrero}, M.~A. and {De Marco}, O.},
        title = "{Analysis of far-UV data of central stars of planetary nebulae: Occurrence and variability of stellar winds}",
      journal = {\aap},
         year = 2013,
        month = may,
       volume = {553},
          eid = {A126},
        pages = {A126},
          doi = {10.1051/0004-6361/201220623},
archivePrefix = {arXiv},
       eprint = {1303.1530},
 primaryClass = {astro-ph.GA},
       adsurl = {https://ui.adsabs.harvard.edu/abs/2013A&A...553A.126G}
}

@ARTICLE{Borkowski1995,
       author = {{Borkowski}, Kazimierz J. and {Harrington}, J. Patrick and {Tsvetanov}, Zlatan I.},
        title = "{Interaction of a Stellar Wind with Clumpy Stellar Ejecta in A30}",
      journal = {\apjl},
         year = 1995,
        month = aug,
       volume = {449},
        pages = {L143},
          doi = {10.1086/309643},
       adsurl = {https://ui.adsabs.harvard.edu/abs/1995ApJ...449L.143B}
}

@ARTICLE{Borkowski1993,
       author = {{Borkowski}, Kazimierz J. and {Harrington}, J.~P. and {Tsvetanov}, Zlatan and {Clegg}, Robin E.~S.},
        title = "{HST Imaging of Hydrogen-poor Ejecta in Abell 30 and Abell 78: Wind-blown Cometary Structures}",
      journal = {\apjl},
         year = 1993,
        month = sep,
       volume = {415},
        pages = {L47},
          doi = {10.1086/187029},
       adsurl = {https://ui.adsabs.harvard.edu/abs/1993ApJ...415L..47B}
}

@ARTICLE{Sabin2022,
       author = {{Sabin}, L. and {G{\'o}mez-Llanos}, V. and {Morisset}, C. and {G{\'o}mez-Gonz{\'a}lez}, V.~M.~A. and {Guerrero}, M.~A. and {Todt}, H. and {Fang}, X.},
        title = "{Catching a grown-up starfish planetary nebula - II. Plasma analysis and central star properties of PC 22}",
      journal = {\mnras},
         year = 2022,
        month = mar,
       volume = {511},
       number = {1},
        pages = {1-19},
          doi = {10.1093/mnras/stab3649},
archivePrefix = {arXiv},
       eprint = {2112.09652},
 primaryClass = {astro-ph.SR},
       adsurl = {https://ui.adsabs.harvard.edu/abs/2022MNRAS.511....1S}
}

@ARTICLE{Gesicki2006,
       author = {{Gesicki}, K. and {Zijlstra}, A.~A. and {Acker}, A. and {G{\'o}rny}, S.~K. and {Gozdziewski}, K. and {Walsh}, J.~R.},
        title = "{Planetary nebulae with emission-line central stars}",
      journal = {\aap},
         year = 2006,
        month = jun,
       volume = {451},
       number = {3},
        pages = {925-935},
          doi = {10.1051/0004-6361:20054192},
archivePrefix = {arXiv},
       eprint = {astro-ph/0601439},
 primaryClass = {astro-ph},
       adsurl = {https://ui.adsabs.harvard.edu/abs/2006A&A...451..925G}
}

@ARTICLE{Guerrero2012,
       author = {{Guerrero}, M.~A. and {Ruiz}, N. and {Hamann}, W. -R. and {Chu}, Y. -H. and {Todt}, H. and {Sch{\"o}nberner}, D. and {Oskinova}, L. and {Gruendl}, R.~A. and {Steffen}, M. and {Blair}, W.~P. and {Toal{\'a}}, J.~A.},
        title = "{Rebirth of X-Ray Emission from the Born-again Planetary Nebula A30}",
      journal = {\apj},
         year = 2012,
        month = aug,
       volume = {755},
       number = {2},
          eid = {129},
        pages = {129},
          doi = {10.1088/0004-637X/755/2/129},
archivePrefix = {arXiv},
       eprint = {1202.4463},
 primaryClass = {astro-ph.GA},
       adsurl = {https://ui.adsabs.harvard.edu/abs/2012ApJ...755..129G}
}

@ARTICLE{Toala2015,
       author = {{Toal{\'a}}, J.~A. and {Guerrero}, M.~A. and {Todt}, H. and {Hamann}, W. -R. and {Chu}, Y. -H. and {Gruendl}, R.~A. and {Sch{\"o}nberner}, D. and {Oskinova}, L.~M. and {Marquez-Lugo}, R.~A. and {Fang}, X. and {Ramos-Larios}, G.},
        title = "{The Born-again Planetary Nebula A78: An X-Ray Twin of A30}",
      journal = {\apj},
         year = 2015,
        month = jan,
       volume = {799},
       number = {1},
          eid = {67},
        pages = {67},
          doi = {10.1088/0004-637X/799/1/67},
archivePrefix = {arXiv},
       eprint = {1411.3837},
 primaryClass = {astro-ph.SR},
       adsurl = {https://ui.adsabs.harvard.edu/abs/2015ApJ...799...67T}
}

@ARTICLE{Keller2014,
       author = {{Keller}, Graziela R. and {Bianchi}, Luciana and {Maciel}, Walter J.},
        title = "{UV spectral analysis of very hot H-deficient [WCE]-type central stars of planetary nebulae: NGC 2867, NGC 5189, NGC 6905, Pb 6 and Sand 3}",
      journal = {\mnras},
         year = 2014,
        month = aug,
       volume = {442},
       number = {2},
        pages = {1379-1395},
          doi = {10.1093/mnras/stu878},
archivePrefix = {arXiv},
       eprint = {1405.6763},
 primaryClass = {astro-ph.SR},
       adsurl = {https://ui.adsabs.harvard.edu/abs/2014MNRAS.442.1379K}
}

@ARTICLE{Toala2019_ngc40,
       author = {{Toal{\'a}}, J.~A. and {Ramos-Larios}, G. and {Guerrero}, M.~A. and {Todt}, H.},
        title = "{Hidden IR structures in NGC 40: signpost of an ancient born-again event}",
      journal = {\mnras},
         year = 2019,
        month = may,
       volume = {485},
       number = {3},
        pages = {3360-3369},
          doi = {10.1093/mnras/stz624},
archivePrefix = {arXiv},
       eprint = {1902.11219},
 primaryClass = {astro-ph.SR},
       adsurl = {https://ui.adsabs.harvard.edu/abs/2019MNRAS.485.3360T}
}

@ARTICLE{GarciaRojas2013,
       author = {{Garc{\'\i}a-Rojas}, Jorge and {Pe{\~n}a}, Miriam and {Morisset}, Christophe and {Delgado-Inglada}, Gloria and {Mesa-Delgado}, Adal and {Ruiz}, Mar{\'\i}a Teresa},
        title = "{Analysis of chemical abundances in planetary nebulae with [WC] central stars. II. Chemical abundances and the abundance discrepancy factor}",
      journal = {\aap},
         year = 2013,
        month = oct,
       volume = {558},
          eid = {A122},
        pages = {A122},
          doi = {10.1051/0004-6361/201322354},
archivePrefix = {arXiv},
       eprint = {1309.1663},
 primaryClass = {astro-ph.GA},
       adsurl = {https://ui.adsabs.harvard.edu/abs/2013A&A...558A.122G}
}

@ARTICLE{Medina2006,
       author = {{Medina}, S. and {Pe{\~n}a}, M. and {Morisset}, C. and {Stasi{\'n}ska}, G.},
        title = "{Galactic Planetary Nebulae with Wolf-Rayet Nuclei III. Kinematical Analysis of a Large Sample of Nebulae}",
      journal = {\rmxaa},
         year = 2006,
        month = apr,
       volume = {42},
        pages = {53-74},
          doi = {10.48550/arXiv.astro-ph/0603651},
archivePrefix = {arXiv},
       eprint = {astro-ph/0603651},
 primaryClass = {astro-ph},
       adsurl = {https://ui.adsabs.harvard.edu/abs/2006RMxAA..42...53M}
}

@ARTICLE{Toala2024WR,
       author = {{Toal{\'a}}, Jes{\'u}s A. and {Todt}, Helge and {Sander}, Andreas A.~C.},
        title = "{Peering into the Wolf-Rayet phenomenon through [WO] and [WC] stars}",
      journal = {\mnras},
         year = 2024,
        month = jun,
       volume = {531},
       number = {2},
        pages = {2422-2432},
          doi = {10.1093/mnras/stae1298},
archivePrefix = {arXiv},
       eprint = {2405.10177},
 primaryClass = {astro-ph.SR},
       adsurl = {https://ui.adsabs.harvard.edu/abs/2024MNRAS.531.2422T}
}

@ARTICLE{Corradi2000,
       author = {{Corradi}, R.~L.~M. and {Sch{\"o}nberner}, D. and {Steffen}, M. and {Perinotto}, M.},
        title = "{A hydrodynamical study of multiple-shell planetaries . I. NGC 2438}",
      journal = {\aap},
         year = 2000,
        month = feb,
       volume = {354},
        pages = {1071-1085},
       adsurl = {https://ui.adsabs.harvard.edu/abs/2000A&A...354.1071C}
}

@ARTICLE{Welch1999,
       author = {{Welch}, C.~A. and {Frank}, A. and {Pipher}, Judith L. and {Forrest}, W.~J. and {Woodward}, Charles E.},
        title = "{[Fe II] Bubbles in the Young Planetary Nebula Hubble 12}",
      journal = {\apjl},
         year = 1999,
        month = sep,
       volume = {522},
       number = {1},
        pages = {L69-L72},
          doi = {10.1086/312206},
       adsurl = {https://ui.adsabs.harvard.edu/abs/1999ApJ...522L..69W}
}

@BOOK{Lamers1999,
       author = {{Lamers}, Henny J.~G.~L.~M. and {Cassinelli}, Joseph P.},
        title = "{Introduction to Stellar Winds}",
         year = 1999,
       adsurl = {https://ui.adsabs.harvard.edu/abs/1999isw..book.....L}
}

@article{paxton2011,
  author = {Paxton, Bill and Bildsten, Lars and Dotter, Aaron and Herwig, Falk and Lesaffre, Pierre and Timmes, F. X.},
  title = {Modules for Experiments in Stellar Astrophysics (MESA)},
  journal = {The Astrophysical Journal Supplement Series},
  year = {2011},
  volume = {192},
  number = {1},
  pages = {3},
  doi = {10.1088/0067-0049/192/1/3}
}

@article{paxton2013,
  author = {Paxton, Bill and Cantiello, Matteo and Arras, Phil and Bildsten, Lars and Brown, Edward F. and Dotter, Aaron and Mankovich, Christopher and Montgomery, Michael H. and Stello, Dennis and Timmes, F. X. and Townsend, Richard H. D.},
  title = {Modules for Experiments in Stellar Astrophysics (MESA): Planets, Oscillations, Rotation, and Massive Stars},
  journal = {The Astrophysical Journal Supplement Series},
  year = {2013},
  volume = {208},
  number = {1},
  pages = {4},
  doi = {10.1088/0067-0049/208/1/4}
}

@article{paxton2015,
  author = {Paxton, Bill and Marchant, Pablo and Schwab, Josiah and Bauer, Evan B. and Bildsten, Lars and Cantiello, Matteo and Dessart, Luc and Farmer, R. and Hu, H. and Langer, N. and Townsend, Richard H. D. and Timmes, F. X.},
  title = {Modules for Experiments in Stellar Astrophysics (MESA): Binaries, Pulsations, and Explosions},
  journal = {The Astrophysical Journal Supplement Series},
  year = {2015},
  volume = {220},
  number = {1},
  pages = {15},
  doi = {10.1088/0067-0049/220/1/15}
}

@article{paxton2018,
  author = {Paxton, Bill and Schwab, Josiah and Bauer, Evan B. and Bildsten, Lars and Blinnikov, Sergei and Duffell, Paul C. and Farmer, R. and Hu, H. and Langer, N. and Townsend, Richard H. D. and Timmes, F. X.},
  title = {Modules for Experiments in Stellar Astrophysics (MESA): Convective Boundaries, Element Diffusion, and Massive Star Explosions},
  journal = {The Astrophysical Journal Supplement Series},
  year = {2018},
  volume = {234},
  number = {2},
  pages = {34},
  doi = {10.3847/1538-4365/aaa5a8}
}

@article{paxton2019,
  author = {Paxton, Bill and Smolec, Radoslaw and Schwab, Josiah and Gautschy, Alfred and Bildsten, Lars and Cantiello, Matteo and Dotter, Aaron and Farmer, R. and Goldberg, Joel A. and Jermyn, Adam S. and Marchant, Pablo and Pedersen, Michael G. and Thompson, Todd A. and Townsend, Richard H. D. and Timmes, F. X.},
  title = {Modules for Experiments in Stellar Astrophysics (MESA): Pulsating Variable Stars, Rotation, Convective Boundaries, and Energy Conservation},
  journal = {The Astrophysical Journal Supplement Series},
  year = {2019},
  volume = {243},
  number = {1},
  pages = {10},
  doi = {10.3847/1538-4365/ab2241}
}

@article{reimers1975,
  author = {Reimers, D.},
  title = {Circumstellar absorption lines and mass loss from red giants},
  journal = {Memoires of the Societe Royale des Sciences de Liege},
  year = {1975},
  volume = {8},
  pages = {369}
}

@ARTICLE{blocker1995,
       author = {{Bl\"{o}cker}, T.},
        title = "{Stellar evolution of low and intermediate-mass stars. I. Mass loss on the AGB and its consequences for stellar evolution.}",
      journal = {\aap},
         year = 1995,
        month = may,
       volume = {297},
        pages = {727},
       adsurl = {https://ui.adsabs.harvard.edu/abs/1995A&A...297..727B}
}

@ARTICLE{Blocker1995_postAGB,
       author = {{Bl\"{o}cker}, T.},
        title = "{Stellar evolution of low- and intermediate-mass stars. II. Post-AGB evolution.}",
      journal = {\aap},
         year = 1995,
        month = jul,
       volume = {299},
        pages = {755},
       adsurl = {https://ui.adsabs.harvard.edu/abs/1995A&A...299..755B}
}

@ARTICLE{millerbertolami2016,
       author = {{Miller Bertolami}, Marcelo Miguel},
        title = "{New models for the evolution of post-asymptotic giant branch stars and central stars of planetary nebulae}",
      journal = {\aap},
         year = 2016,
        month = apr,
       volume = {588},
          eid = {A25},
        pages = {A25},
          doi = {10.1051/0004-6361/201526577},
archivePrefix = {arXiv},
       eprint = {1410.1679},
 primaryClass = {astro-ph.SR},
       adsurl = {https://ui.adsabs.harvard.edu/abs/2016A&A...588A..25M}
}

@ARTICLEHerwig{Herwig2001,
       author = {{Herwig}, Falk},
        title = "{The Evolutionary Timescale of Sakurai's Object: A Test of Convection Theory?}",
      journal = {\apjl},
         year = 2001,
        month = jun,
       volume = {554},
       number = {1},
        pages = {L71-L74},
          doi = {10.1086/320911},
archivePrefix = {arXiv},
       eprint = {astro-ph/0103002},
 primaryClass = {astro-ph},
       adsurl = {https://ui.adsabs.harvard.edu/abs/2001ApJ...554L..71H}
}

@ARTICLE{Jermyn2023,
       author = {{Jermyn}, Adam S. and {Bauer}, Evan B. and {Schwab}, Josiah and {Farmer}, R. and {Ball}, Warrick H. and {Bellinger}, Earl P. and {Dotter}, Aaron and {Joyce}, Meridith and {Marchant}, Pablo and {Mombarg}, Joey S.~G. and {Wolf}, William M. and {Sunny Wong}, Tin Long and {Cinquegrana}, Giulia C. and {Farrell}, Eoin and {Smolec}, R. and {Thoul}, Anne and {Cantiello}, Matteo and {Herwig}, Falk and {Toloza}, Odette and {Bildsten}, Lars and {Townsend}, Richard H.~D. and {Timmes}, F.~X.},
        title = "{Modules for Experiments in Stellar Astrophysics (MESA): Time-dependent Convection, Energy Conservation, Automatic Differentiation, and Infrastructure}",
      journal = {\apjs},
         year = 2023,
        month = mar,
       volume = {265},
       number = {1},
          eid = {15},
        pages = {15},
          doi = {10.3847/1538-4365/acae8d},
archivePrefix = {arXiv},
       eprint = {2208.03651},
 primaryClass = {astro-ph.SR},
       adsurl = {https://ui.adsabs.harvard.edu/abs/2023ApJS..265...15J}
}

@article{Frew2010,
  author  = {Frew, D. J. and Parker, Q. A.},
  title   = {Planetary Nebulae: Observational Properties, Mimics and Diagnostics},
  journal = {Publications of the Astronomical Society of Australia},
  year    = {2010},
  volume  = {27},
  number  = {2},
  pages   = {129--148},
  doi     = {10.1071/AS09040}
}

@article{Frew2016,
  author  = {Frew, D. J. and Parker, Q. A. and Boji\v{c}i\'{c}, I. S.},
  title   = {The H$\alpha$ surface brightness-–radius relation: a robust statistical distance indicator for planetary nebulae},
  journal = {Monthly Notices of the Royal Astronomical Society},
  year    = {2016},
  volume  = {455},
  number  = {2},
  pages   = {1459--1488},
  doi     = {10.1093/mnras/stv2189}
}

@ARTICLE{Ferland2017,
       author = {{Ferland}, G.~J. and {Chatzikos}, M. and {Guzm{\'a}n}, F. and {Lykins}, M.~L. and {van Hoof}, P.~A.~M. and {Williams}, R.~J.~R. and {Abel}, N.~P. and {Badnell}, N.~R. and {Keenan}, F.~P. and {Porter}, R.~L. and {Stancil}, P.~C.},
        title = "{The 2017 Release Cloudy}",
      journal = {\rmxaa},
         year = 2017,
        month = oct,
       volume = {53},
        pages = {385-438},
          doi = {10.48550/arXiv.1705.10877},
archivePrefix = {arXiv},
       eprint = {1705.10877},
 primaryClass = {astro-ph.GA},
       adsurl = {https://ui.adsabs.harvard.edu/abs/2017RMxAA..53..385F}
}

@ARTICLE{Kwok1978,
       author = {{Kwok}, S. and {Purton}, C.~R. and {Fitzgerald}, P.~M.},
        title = "{On the origin of planetary nebulae.}",
      journal = {\apjl},
         year = 1978,
        month = feb,
       volume = {219},
        pages = {L125-L127},
          doi = {10.1086/182621},
       adsurl = {https://ui.adsabs.harvard.edu/abs/1978ApJ...219L.125K}
}

@ARTICLE{Orozco-Duarte2026,
       author = {{Orozco-Duarte}, Rogelio and {Toal{\'a}}, Jes{\'u}s A. and {Arthur}, S. Jane and {Rodr{\'\i}guez-Gonz{\'a}lez}, Janis B. and {Conmy}, Luke and {Kuiper}, Rolf},
        title = "{Formation and X-ray emission from hot bubbles in planetary nebulae ─ III. The impact of [Wolf─Rayet]-type winds}",
      journal = {\mnras},
         year = 2026,
        month = mar,
       volume = {546},
       number = {3},
          eid = {stag179},
        pages = {stag179},
          doi = {10.1093/mnras/stag179},
archivePrefix = {arXiv},
       eprint = {2601.16100},
 primaryClass = {astro-ph.SR},
       adsurl = {https://ui.adsabs.harvard.edu/abs/2026MNRAS.546ag179O}
}

@ARTICLE{Schonberner2005,
       author = {{Sch{\"o}nberner}, D. and {Jacob}, R. and {Steffen}, M. and {Perinotto}, M. and {Corradi}, R.~L.~M. and {Acker}, A.},
        title = "{The evolution of planetary nebulae.  II. Circumstellar environment and expansion properties}",
      journal = {\aap},
         year = 2005,
        month = mar,
       volume = {431},
       number = {3},
        pages = {963-978},
          doi = {10.1051/0004-6361:20041669},
       adsurl = {https://ui.adsabs.harvard.edu/abs/2005A&A...431..963S}
}

@ARTICLE{Liu2025,
       author = {{Liu}, Zhongyang and {Jeffery}, C. Simon and {Zhang}, Xianfei and {Bi}, Shaolan and {Li}, Tanda},
        title = "{Theoretical Models for the Late Thermal Pulse in Post-asymptotic Giant Branch Stars: The Case of DY Cen}",
      journal = {\apj},
         year = 2025,
        month = dec,
       volume = {995},
       number = {1},
          eid = {114},
        pages = {114},
          doi = {10.3847/1538-4357/ae182d},
archivePrefix = {arXiv},
       eprint = {2510.26250},
 primaryClass = {astro-ph.SR},
       adsurl = {https://ui.adsabs.harvard.edu/abs/2025ApJ...995..114L}
}

@ARTICLE{Wagstaff2020M,
       author = {{Wagstaff}, G. and {Miller Bertolami}, M.~M. and {Weiss}, A.},
        title = "{Impact of convective boundary mixing on the TP-AGB}",
      journal = {\mnras},
         year = 2020,
        month = apr,
       volume = {493},
       number = {4},
        pages = {4748-4762},
          doi = {10.1093/mnras/staa362},
archivePrefix = {arXiv},
       eprint = {2002.01860},
 primaryClass = {astro-ph.SR},
       adsurl = {https://ui.adsabs.harvard.edu/abs/2020MNRAS.493.4748W}
}
%\begin{thebibliography}{99}
%\end{thebibliography}

%%%%%%%%%%%%%%%%%%%%%%%%%%%%%%%%%%%%
\appendix

\section{The AGB phase evolution}
\label{app:AGB}

\begin{figure}
    \centering
    \includegraphics[width=1.0\linewidth]{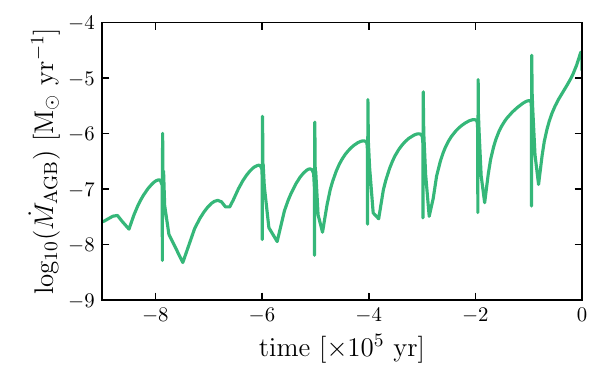}\\[-22pt]
    \includegraphics[width=1.0\linewidth]{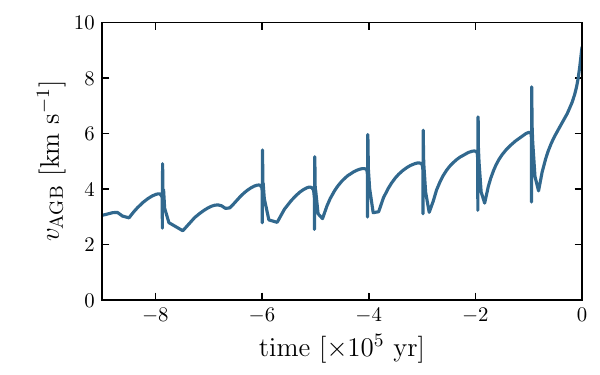}
    \caption{Time evolution of the stellar wind mass-loss rate (top) and terminal velocity (bottom) during the last $9\times10^{5}$ yr of evolution of the AGB phase.}
    \label{fig:AGB}
\end{figure}

The first evolutionary stage modelled by the {\sc pluto} radiation-hydrodynamic simulations is that of the AGB phase. The wind properties are injected into the grid, using the mass-loss rate ($\dot{M}_\mathrm{AGB}$) directly obtained from {\sc mesa}. However, the stellar wind velocity during this phase needs to be computed. This was done by adopting the empirical relation derived by \citet{Verbena2011}
\begin{equation}
    \varv_\mathrm{AGB} = 0.05 \left( \frac{L}{\mathrm{L}_\odot} \cdot \frac{\mathrm{M}_\odot}{M} \right)^{0.57}~\mathrm{km~s}^{-1}, 
\end{equation}
\noindent where $L$ and $M$ are the luminosity and mass of the evolving star. Fig.~\ref{fig:AGB}
shows the evolution of $\dot{M}_\mathrm{AGB}$ and $\varv_\mathrm{AGB}$ for the last $9\times10^{5}$~yr of evolution during the TP-AGB phase. No ionisation flux is included during this phase.

\begin{figure}
    \centering
    \includegraphics[width=1.0\linewidth]{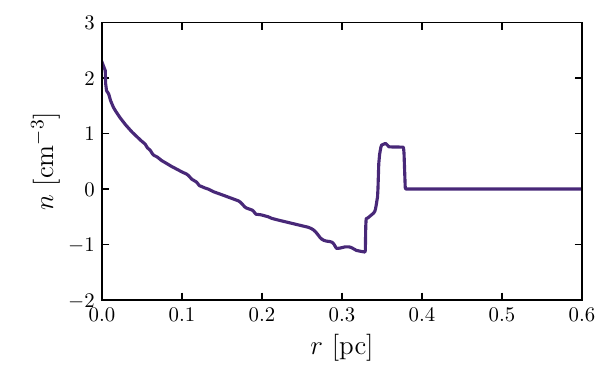}
    \caption{Number density profile of the CSM at the end of the AGB phase. The ISM is the region with a number density of 1 cm$^{-3}$ beyond 0.4 pc.}
    \label{fig:AGB_density}
\end{figure}

Fig.~\ref{fig:AGB_density} shows the density configuration of the CSM just before the onset of the post-AGB phase (the PN phase). As expected, the density structure has a $\sim r^{-2}$ profile. The figure depicts a jump in density which marks the interaction of the AGB material with the ISM. The total mass ejected during the TP-AGB phase is about 1.02 M$_\odot$.

\section{Evolution of a standard PN}
\label{app:PN}

\begin{figure*}
    \includegraphics[width=.95\linewidth]{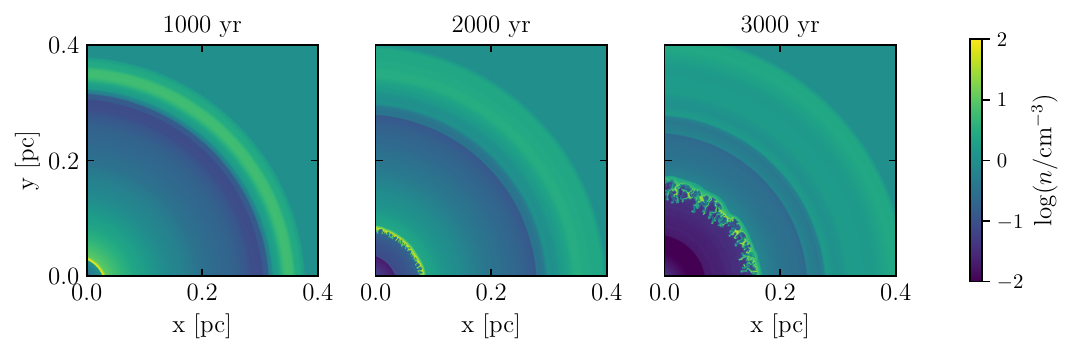}\\[-24pt]
    \includegraphics[width=.95\linewidth]{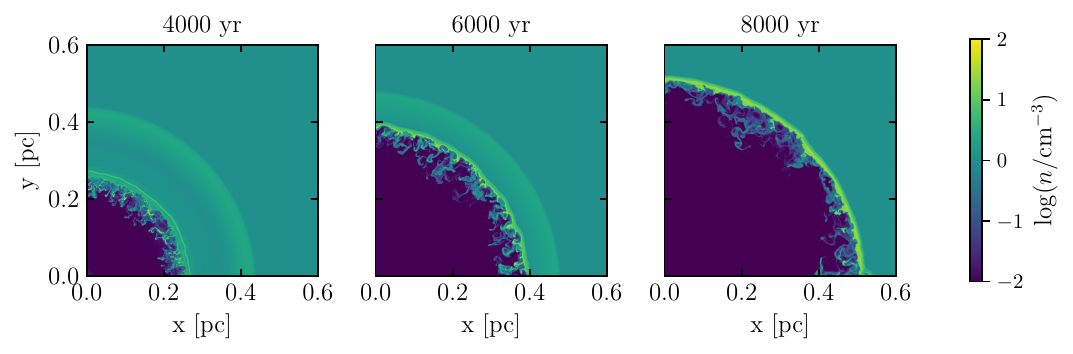}
    \caption{ Number density $n$ maps showing the temporal evolution of the CSM around the nLTP model. $t=0$ yr is the onset of the post-AGB phase.}
    \label{fig:density_nLTP}
\end{figure*}

\begin{figure*}
   \includegraphics[width=.95\linewidth]{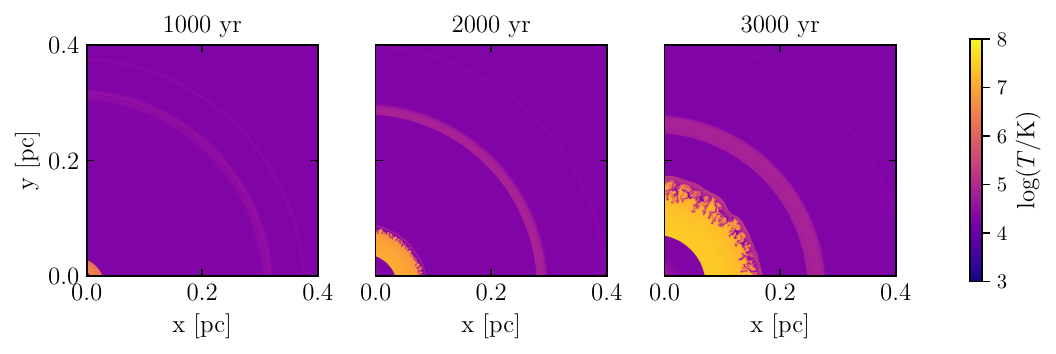}\\[-24pt]
   \includegraphics[width=.95\linewidth]{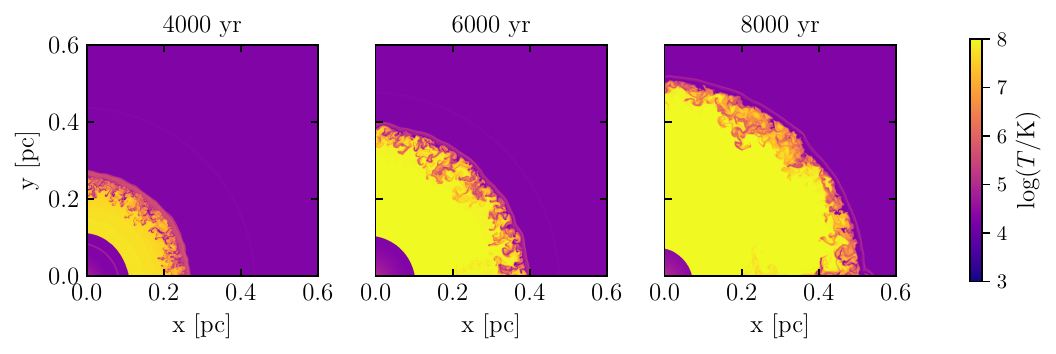}
   \caption{Gas temperature $T$ maps showing the temporal evolution of the CSM around the nLTP model. $t=0$ yr is the onset of the post-AGB phase.}
   \label{fig:temperature_nLTP}
\end{figure*}

This appendix presents the formation and evolution of a typical PN that does not undergo a born-again event. Fig.~\ref{fig:density_nLTP} and \ref{fig:temperature_nLTP} illustrate the temporal evolution of the gas number density $n$ and temperature $T$ obtained from simulations based on the stellar evolution model shown in the bottom panel of Fig.~\ref{fig:HRD2}. The corresponding evolution of the stellar wind parameters are presented as dashed lines in Fig.~\ref{fig:postAGB}.

Fig.~\ref{fig:density_nLTP} and \ref{fig:temperature_nLTP} show the well-known formation of a hot bubble driven by the interaction between the fast post-AGB wind and the slower material ejected during the AGB phase (see references in Section~\ref{sec:intro}). As the fast wind impacts the circumstellar medium, hydrodynamical instabilities naturally develop at the wind-wind interaction interface, promoting mixing between the shocked stellar wind and the nebular gas.

After 8000 yr of evolution during the post-AGB phase, the dense rim reached a distance of about 0.5 pc.

%%%%%%%%%%%%%%%%%%%%%%%%%%%%%%%%%%%%

\end{document}